\documentclass[a4paper]{article}
\usepackage{amsmath,amssymb,theorem,fullpage,mathtools}
\usepackage[shortcuts]{extdash}
\usepackage{hyperref}
\usepackage[nosort]{cite}

\let\frac\undefined

\allowdisplaybreaks

\numberwithin{equation}{section}

\def\Maketitle{{\def\newpage{}\maketitle}}

\def\eq#1$$#2$${\begin{equation#1}#2\end{equation#1}}
\long\def\subeq#1{\begin{subequations}#1\end{subequations}}
\def\Split$$#1$${\begin{split}#1\end{split}}
\def\Align#1$$#2$${\begin{align#1}#2\end{align#1}}
\def\AlignAt#1$$#2$${\begin{alignat}{#1}#2\end{alignat}}
\def\Aligned#1{\begin{aligned}#1\end{aligned}}
\def\Gather#1$$#2$${\begin{gather#1}#2\end{gather#1}}
\def\Gathered#1{\begin{gathered}#1\end{gathered}}
\def\Multline#1$$#2$${\begin{multline#1}#2\end{multline#1}}

\def\?{\notag}
\def\d{\partial}
\def\bd{\bar\partial}

\def\Im{\mathop{\rm Im}\nolimits}

\def\Res{\mathop{\rm Res}}

\def\const{\mathop{\rm const}\nolimits}
\def\cA{{\cal A}}

\def\cC{{\cal C}}

\def\cF{{\cal F}}
\def\bcF{{\cal\bar F}}

\def\cH{{\cal H}}

\def\cO{{\cal O}}
\def\cR{{\cal R}}

\def\sh{\mathop{\rm sh}\nolimits}
\def\ch{\mathop{\rm ch}\nolimits}
\def\th{\mathop{\rm th}\nolimits}

\def\tg{\mathop{\rm tg}\nolimits}
\def\ctg{\mathop{\rm ctg}\nolimits}
\def\lcolon{\mathopen{\,:\,}}
\def\rcolon{\mathclose{\,:\,}}
\def\C{{\mathbb C}}

\def\Z{{\mathbb Z}}
\def\bbD{{\mathbb D}}
\def\bbDodd{{\mathbb D}^{\rm odd}}
\def\bbDeven{{\mathbb D}^{\rm even}}

\def\bfd{{\bf d}}

\def\bfD{{\bf D}}

\def\ft{{\mathfrak t}}
\def\tft{{\tilde{\mathfrak t}}}

\def\e{{\rm e}}
\def\i{{\rm i}}

\def\tB{{\tilde B}}

\def\tS{{\tilde S}}
\def\tq{{\tilde q}}
\def\tQ{{\tilde Q}}

\def\tSigma{{\tilde\Sigma}}

\def\tdelta{{\tilde\delta}}

\def\bT{{\bar T}}
\def\bz{{\bar z}}

\def\bc{{\bar c}}
\def\bh{{\bar h}}
\def\bcA{{\bar{\cal A}}}

\def\teta{{\tilde\eta}}

\makeatletter

\def\section{\@startsection{section}{1}{\z@}%
                                   {-3.5ex \@plus -1ex \@minus -.2ex}%
                                   {2.3ex \@plus.2ex}%
                                   {\normalfont\normalsize\bfseries}}
\def\subsection{\@startsection{subsection}{2}{\z@}%
                                     {-3.25ex\@plus -1ex \@minus -.2ex}%
                                     {1.5ex \@plus .2ex}%
                                     {\normalfont\normalsize\bfseries\itshape}}
\def\@seccntformat#1{\csname the#1\endcsname.~~}
\long\def\@makecaption#1#2{%
  \vskip\abovecaptionskip
  \sbox\@tempboxa{\small#1. #2}%
  \ifdim \wd\@tempboxa >0.9\hsize
  {\leftskip=0.05\hsize\rightskip=0.05\hsize\relax\small
    #1. #2\par}
  \else
    \global \@minipagefalse
    \hb@xt@\hsize{\hfil\box\@tempboxa\hfil}%
  \fi
  \vskip\belowcaptionskip}
\def\Appendix{\appendix
  \def\@seccntformat##1{Appendix~\csname the##1\endcsname.~~}}

\let\over\@@over
\let\atop\@@atop
\let\above\@@above
\let\overwithdelims\@@overwithdelims
\let\atopwithdelims\@@atopwithdelims
\let\abovewithdelims\@@abovewithdelims

\makeatother

\begin{document}

\title{Form factors of descendant operators:\\ Reduction to perturbed $M(2,2s+1)$ models}
\author{Michael Lashkevich${}^{1-3}$ and Yaroslav Pugai${}^{1,2}$\\[\medskipamount]
\parbox[t]{0.9\textwidth}{\normalsize\it\raggedright
${}^1$~Landau Institute for Theoretical Physics, 142432 Chernogolovka, Russia\medspace%
\footnote{Mailing address.}
\\
${}^2$~Moscow Institute of Physics and Technology, 141707 Dolgoprudny, Russia\\
${}^3$~Kharkevich Institute for Information Transmission Problems, 19 Bolshoy Karetny per., 127994 Moscow, Russia}
}
\date{}

\Maketitle

\begin{abstract}

In the framework of the algebraic approach to form factors in two\-/dimensional integrable models of quantum field theory we consider the reduction of the sine\-/Gordon model to the $\Phi_{13}$\=/perturbation of minimal conformal models of the $M(2,2s+1)$ series. We find in an algebraic form the condition of compatibility of local operators with the reduction. We propose a construction that make it possible to obtain reduction compatible local operators in terms of screening currents. As an application we obtain exact multiparticle form factors for the compatible with the reduction conserved currents $T_{\pm2k}$, $\Theta_{\pm(2k-2)}$, which correspond to the spin $\pm(2k-1)$ integrals of motion, for any positive integer~$k$. Furthermore, we obtain all form factors of the operators $T_{2k}T_{-2l}$, which generalize the famous $T\bar T$ operator. The construction is analytic in the $s$ parameter and, therefore, makes sense in the sine\-/Gordon theory.

\end{abstract}

\section{Introduction}

We continue studying the algebraic approach to form factors of descendant operators in two\-/dimensional massive integrable models of quantum field theory proposed in~\cite{Feigin:2008hs} and developed in~\cite{Alekseev:2009ik,Lashkevich:2013mca,Lashkevich:2013yja,Lashkevich:2014rua}.

Most of massive integrable field theories in two dimensions can be considered as relevant perturbations of conformal models\cite{Zamolodchikov:1987jf,Zamolodchikov:1989zs}. The analysis of correlation functions of conformal field theories is simplified by the fact that they possess an infinite-dimensional non-commutative symmetry algebra, the product of two chiral Virasoro algebras in the simplest case~\cite{Belavin:1984vu}. The space of local operators splits into a sum of subspaces, each of which is a product of two infinite-dimensional irreducible representations of the right and left chiral algebras. Within each term the two chiralities decouple, which results in splitting the correlation function into a sum of products of purely algebraic objects: conformal blocks. Methods of representation theory fail in perturbed models, and analysis of local operators is more involved in this case. However, there are methods to study the long-range and short-range assymptotics of correlation functions.

The short-range (ultraviolet) expansion of correlation functions can be effectively developed for any conformal model perturbed by a relevant operator by using the method of conformal perturbation theory~\cite{Zamolodchikov:1990bk}. In a perturbed model local operators can be considered as perturbations of those at the conformal point, but need admixing other operators for renormalization. As soon as vacuum expectation values of local operators and structure constants of the operator algebra at the conformal point are known, the conformal perturbation theory can be effectively applied to analysis of primary~\cite{Zamolodchikov:1990bk} as well as descendant operators~\cite{Belavin:2005xg, Fateev:2006js,Fateev:2009kp}. Of course, special attention should be paid to the resonance phenomenon \cite{Zamolodchikov:1990bk}, which takes place at thresholds of ultraviolet divergences and is responsible for logarithmic contributions to the short-range assymptotics of correlation functions.

The long-range (infrared) expansion can be constructed in terms of form factors, which are matrix elements of local operators with respect to the eigenvectors of the Hamiltonian. In the case of massive integrable models, form factors are solutions to a system of linear difference equations called form factor axioms\cite{Smirnov:1992vz}. These equations are bootstrap equations, i.e.\ they follow from the general requirements of consistency and integrability. They involve the exact spectrum of the model and the exact $S$ matrix, which are supposed to be known. Every solution to the equations provides a set of form factors, which define a local (or quasi-local) operator. Once the solution, which corresponds to a given operator in the sense of conformal perturbation theory, has been found, it is possible to obtain a reasonable numeric approximation to correlation functions by interpolating between the infrared and ultraviolet expansions\cite{Zamolodchikov:1990bk,Belavin:2003pu,Belavin:2005xg}.

General solutions of the form factor equations for a number of models were found by F.~Smirnov~(see~\cite{Smirnov:1992vz} and references therein). An important step in developing a systematic approach to finding form factors was done by S.~Lukyanov~\cite{Lukyanov:1993pn,Lukyanov:1997bp}, who modified the free field technique known from conformal field theory~\cite{Dotsenko:1984nm} to obtaining form factors. Nevertheless, the problem of identifying these solutions with some particular operators known from the conformal perturbation theory has not yet been solved. For most of studied integrable models form factors are known for a set of the simplest local operators, usually for primary operators of the underlying conformal models. However, great efforts were made to find form factors of other operators, which are descendant operators from the point of view of conformal field theory~\cite{Cardy:1990pc,Babelon:1996sk,Babujian:2002fi,Babujian:2003za,Delfino:2004vc,Delfino:2005wi,Delfino:2006te, Fateev:2006js,Fateev:2009kp}. In particular, counting procedures were invented, which reproduce the characters of representations of the Virasoro algebra even outside the conformal points~\cite{Koubek:1993ke,Smirnov:1995jp,Jimbo:2003ne,Delfino:2007bt}. Recently, a novel approach was proposed~\cite{Boos:2006mq,Jimbo:2010jv,Jimbo:2011bc}, which allows identifying some descendant operators (modulo commutators with integrals of motion) by means of a fermion algebra.

An approach to solving form factor equations in theories with diagonal $S$\=/matrix was proposed in~\cite{Feigin:2008hs}. It was initially inspired by Lukyanov's free field representation, but incorporated naturally form factors for descendant operators, resulting in formulas similar to those of~\cite{Babujian:2002fi}. In our previous publications we have shown this approach to be applicable to studying descendant operators that contain both right and left chiralities simultaneously\cite{Lashkevich:2013mca,Lashkevich:2013yja,Lashkevich:2014rua}, but the method was being developed for studying boson theories, like the sinh\-/Gordon model or the breather sector of the sine\-/Gordon model, for generic values of coupling constants. Here we consider the $\Phi_{13}$ perturbation of minimal conformal models of the so called $M(2,2s+1)$ series. One of the ways to obtain form factors of local operators in this theory is the reduction of the sine\-/Gordon theory\cite{LeClair:1989wy,Smirnov:1990vm,Reshetikhin:1989qg,Eguchi:1989dq}. In the framework of the reduction procedure local operators in a perturbed minimal model coincide with particular quasi\-/local operators in the sine\-/Gordon model. These particular operators must satisfy some compatibility conditions, which look as invariance of all their form factors with respect to some symmetries. These conditions look very complicated. We will see that for the perturbed $M(2,2s+1)$ minimal models, the compatibility condition radically simplifies in the framework of the algebraic approach of~\cite{Feigin:2008hs}. Moreover, we develop a constructive way to find operators compatible with the reduction. As a particular example it is explained how to define compatible with the reduction conserved currents and their products. An important feature of the resulting form factors is that they, being analytic in the model parameter, make sense in the sine\-/Gordon model independently of the reduction procedure.

The sine\-/Gordon model is a theory of a real scalar field $\varphi(x)$ with the action
\eq$$
S[\varphi]=\int d^2x\,\left({(\d_\nu\varphi)^2\over16\pi}+2\mu\cos\beta\varphi\right).
\label{sG-action}
$$
It will be convenient to use the parameter
\eq$$
r={1\over1-\beta^2}.
\label{r-def}
$$
The spectrum of the theory consists of a pair of topological solitons of mass $M\sim \mu^{1/(2-2\beta^2)}$ and a set of bound states called breathers with masses
\eq$$
m_\nu=2M\sin{\pi(r-1)\nu\over2},
\qquad
\nu=1,2,\ldots,\nu_{\text{max}}\le{1\over r-1}.
\label{mnu-def}
$$
At rational values of $r$, $r={p'\over p'-p}$, where $p,p'$ is a pair of coprime integers such that $p'>p\ge2$, the sine\-/Gordon model admits a reduction of the space of states. The resulted theory coincides with a perturbation of the $M(p,p')$ minimal conformal model\cite{Smirnov:1990vm}. The $M(p,p')$ minimal model\cite{Belavin:1984vu} is the rational model of the conformal field theory, whose chiral algebra is the pure Virasoro algebra with the central charge
$$
c=1-{6(p'-p)^2\over pp'}.
$$
The model is known to possess a set of primary fields $\Phi_{mn}(x)$, $m=1,2,\ldots,p-1$, $n=1,2,\ldots,p'-1$ with conformal dimensions given by the Kac formula
$$
\Delta_{mn}={(p'm-pn)^2-(p'-p)^2\over4pp'}.
$$
The operators $\Phi_{mn}$ and $\Phi_{p-m,p'-n}$ have the same dimension and coincide. These fields are obtained by means of the reduction procedure from the exponential fields $\e^{\i\alpha_{mn}\varphi(x)}$, where
\eq$$
\alpha_{mn}={1-m\over2}\beta^{-1}-{1-n\over2}\beta.
$$
Note that the same operators $\Phi_{mn}$ is obtained from two different exponential fields $\e^{\i\alpha_{mn}\varphi(x)}$ and $\e^{\i\alpha_{p-m,p'-n}\varphi(x)}$.

Let $S_{M(p,p')}$ be the formal action of the $M(p,p')$ model. The formal action of the theory we are interested in is
\eq$$
S=S_{M(p,p')}-\lambda\int d^2x\,\Phi_{13}(x),
\label{S-min-pert}
$$
where $\lambda\sim\mu^2$ is a coupling constant. The word `formal' means that the explicit form of the action is unknown, but the expression (\ref{S-min-pert}) defines the perturbation theory for correlation functions as soon as exact correlation functions of the conformal model are known.

We shall consider the particular series $p=2$, $p'=2s+1$ with $s=2,3,\ldots\ $ This is the so called `ribbon' series, where the set of primary operators is given by $\Phi_{1n}(x)$, $n=1,\ldots,2s$. Since $\Phi_{1n}(x)=\Phi_{1,2s+1-n}(x)$, we can consider a half of these operators, say those with $n=1,\ldots,s$ or, better, those with odd values of~$n$: $n=1,3,\ldots,2s-1$.

The `ribbon' series of perturbed minimal models is related to the sine\-/Gordon model with
\eq$$
r={2s+1\over2s-1},
\qquad s=2,3,\ldots
\label{rs-values}
$$
For this series the reduction rule is made in two steps:

First, soliton states should be removed: we consider the subspace $\cH_{\text{br}}$ generated by breathers in the whole space of states $\cH$ of the sine\-/Gordon model. It means that the matrix elements of local operators that only contain breather states are the only matrix elements of essence. From the point of view of the space of operators, it amounts to factorization of this space: operators with the same breather form factors are equivalent. It radically simplifies matters, and the algebraic approach of~\cite{Feigin:2008hs} can be applied in the breather sector. In this sector the operators $\e^{\i\alpha_{1n}\varphi}$ and $\e^{\i\alpha_{1,2s+1-n}\varphi}$ coincide up to a $c$\=/number factor.

Second, the first breather ($\nu=1$) is required to coincide with the $(2s-2)$th breather in the reduced model. Moreover, it follows from this requirement that each breather number $\nu$ coincides with the breather number $2s-1-\nu$. This means that the space of states of the minimal model is the factor\-/space $\cH_{\text{br}}/\mathord\sim$, where $\sim$ is the equivalence relation realizing this identification. In other words, a local operator is compatible with the reduction, if every matrix element that contains an instance of the first breather is equal to the matrix element, where this instance is substituted by the $(2s-2)$th breather of the same rapidity. This requirement looks very complicated, since it amounts to an infinite set of nontrivial equations for analytic functions. A treatable formulation of this requirement is just the problem addressed in the present paper.

The last requirement needs to be explained in more detail. According to~(\ref{mnu-def}) the highest breather is that with $\nu=\nu_{\text{max}}=s-1$. In the sine\-/Gordon model this restriction is explained as follows. The $\nu$th breather can be considered as a bound state of $\nu$ first breathers. It is more convenient to consider every $(\nu+1)$th breather as a bound state of a $\nu$th and a first ones. The $S$ matrix of one $\nu$th and one first breathers has the form
\eq$$
S_{\nu1}(\theta)
={\th{1\over2}\left(\theta+{\i\pi(r-1)\over2}(\nu-1)\right)\th{1\over2}\left(\theta+{\i\pi(r-1)\over2}(\nu+1)\right)
  \over\th{1\over2}\left(\theta-{\i\pi(r-1)\over2}(\nu-1)\right)\th{1\over2}\left(\theta-{\i\pi(r-1)\over2}(\nu+1)\right)}.
\label{S11-def}
$$
Here $\theta=\theta_1-\theta_2$ is a difference of the rapidities of two particles. The rapidities of particles are defined in terms of their on-shell momenta: $p_1=(m_\nu\ch\theta_1,m_\nu\sh\theta_1)$ and $p_2=(m_1\ch\theta_2,m_1\sh\theta_2)$.

The function $S_{\nu1}(\theta)$ possesses poles at the points $\theta={\i\pi(r-1)\over2}(\nu-1)$ and  $\theta={\i\pi(r-1)\over2}(\nu+1)$. Each of them can produce a bound state. The first one produces the $(\nu-1)$th breather as a bound state, and we will not be interested in it. The last one produces the $(\nu+1)$th breather, if
\eq$$
(\Gamma^{(\nu)})^2\equiv-\i\Res_{\theta={\i\pi(r-1)\over2}(\nu+1)}S_{\nu1}(\theta)>0.
\label{Gamma(nu)-def}
$$
In fact,
\eq$$
(\Gamma^{(\nu)})^2=2\tg{\pi(r-1)\nu\over2}\tg{\pi(r-1)(\nu+1)\over2}\ctg{\pi(r-1)\over2}
\label{Gamma(nu)2-explicit}
$$
is finite and positive for $\nu+1<1/(r-1)$, which provides the rule~(\ref{mnu-def}). For $1/(r-1)<\nu+1<1+1/(r-1)$ the values of $(\Gamma^{(\nu)})^2$ are negative, and for $\nu+1>2+1/(r-1)$ they become positive again. In a unitary theory an excitation must be excluded from the spectrum, if the corresponding $\Gamma^2<0$, but the $M(2,2s+1)$ theory is known to be nonunitary. We may say that the sine\-/Gordon model stripped off solitons admits a nonunitary extension. For the special values (\ref{rs-values}), due to the evident identity $m_\nu=m_{2s-1-\nu}$ the particles `after the unitarity horizon' ($\nu\ge s+1$) have the same masses as those `before the horizon' ($\nu\le s$). Identification of particles of the same mass defines a perturbed minimal model in the $S$\=/matrix bootstrap approach as well as identification of local operators of the same dimension defines a minimal model in the conformal bootstrap approach. In the present paper we explain how to find operators, defined in terms of their exact form factors, compatible with this identification.

The paper is organized as follows. In Sect.~\ref{ff-sec} we recall the free field realization of~\cite{Feigin:2008hs} and formulate the main problem in terms of the algebraic approach. In Sect.~\ref{scralg-sec} the algebra generated by Laurent components of the screening currents is described and the definition of screening operators is recalled. Partially, this algebra was described in~\cite{Lashkevich:2013yja,Lashkevich:2014rua}, but here we complete it with the `antiscreening' currents. In Sect.~\ref{tau-sec} a new object is introduced, the $\tau^{(\nu)}(z)$ currents, which makes it possible to rewrite the usual particle creating currents as a kind of commutators of $\tau^{(\nu)}(z)$ with a screening operator. A simple structure of these currents allows us to essentially simplify the problem of compatibility with the reduction in Sect.~\ref{compatcond-sec}. In Sect.~\ref{locop-comp-sec} a method of constructing operators compatible with the reduction is proposed. The formula (\ref{JR-def}) for form factors of such operators is the main result of the paper. Creation of local operators compatible with the reduction by means of `antiscreening' currents is described in Sect.~\ref{S+-sec}. In Sect.~\ref{conscur-sec} the general formula (\ref{JR-def}) is applied to obtain form factors of conserved currents and their products compatible with the reduction.

\section{Free field realization of form factors}
\label{ff-sec}

Let $\cO(x)$ be a local operator, $|\nu_1\theta_1,\ldots,\nu_N\theta_N\rangle$ be an eigenstate (defined as an in\-/state) with $N$ particles with the internal states labeled by $\nu_1,\ldots,\nu_N$ and the rapidities $\theta_1<\cdots<\theta_N$. Then the matrix elements
\Multline*$$
\langle\nu_{k+1}\theta_{k+1},\ldots,\nu_N\theta_N|\cO(0)|\nu_1\theta_1,\ldots,\nu_k\theta_k\rangle
\\
=F_\cO(\theta_1,\ldots,\theta_k,\theta_N+\i\pi,\ldots,\theta_{k+1}+\i\pi)_{\nu_1\ldots\nu_k\nu'_{k+1}\ldots\nu'_N}
\prod^N_{i=k+1}C^{\nu'_i\nu_i},
$$
where $C^{\nu'\nu}$ is the charge conjugation matrix, define analytic functions $F_\cO(\theta_1,\ldots,\theta_N)_{\nu_1\ldots\nu_N}$. These functions are called form factors of the local operator $\cO(x)$, and the full set of form factors uniquely defines an operator.

Since we are considering the breather sector of the sine\-/Gordon model, we assume that $\nu_i$ is the number of the breather, while the charge conjugation matrix is the unit matrix. Moreover, we may restrict ourselves by the first breather only, since form factors with higher breathers are obtained by the fusion procedure (see below). If all $\nu_i=1$, we will omit the subscripts. The first breather form factors of local operators has the form
\eq$$
F_\cO(\theta_1,\ldots,\theta_N)
=\rho^NJ_\cO(\e^{\theta_1},\ldots,\e^{\theta_N})\prod^N_{i<j}R(\theta_i-\theta_j),
\label{FJ-rel}
$$
where $\rho$ is a constant and $R(\theta)$ is an $\cO$\=/independent function. Explicitly they are produced in Appendix~\ref{reference-appendix}. The functions $J_\cO(x_1,\ldots,x_N)$ are rational symmetric functions. We will often use the shorthand notation $X=(x_1,\ldots,x_N)$ for the variables and $J_\cO(X)$ for these function. Evaluation of form factors reduces to evaluation of $J$\=/functions.

Let us recall the construction of the $J$\=/functions by means of the free field approach of~\cite{Feigin:2008hs}. Consider the Heisenberg algebra generated by the elements $\d_a$, $\hat a$, $d^\pm_k$ ($k\in\Z\setminus\{0\}$) with the commutation relations
\eq$$
[\d_a,\hat a]=1,
\qquad
[d^\pm_k,d^\mp_l]=kA^\pm_k\delta_{k+l,0},
\label{dpm-commut}
$$
where
\eq$$
A^\pm_k=(\pm)^k(q^{k/2}-q^{-k/2})(q^{k/2}-(-)^kq^{-k/2}).
\label{Apm-def}
$$
All other commutators vanish. Define the vacuum vectors
\eq$$
\Aligned{
&{}_a\langle1|\hat a={}_a\langle1|a,
\qquad
&&\hat a|1\rangle_a=a|1\rangle_a,
\\
&{}_a\langle1|d^\pm_{-k}=0,
\qquad
&&d^\pm_k|1\rangle_a=0
\qquad
(k>0).
}\label{vac-def}
$$
The action of the Heisenberg algebra on the vacuum vectors ${}_a\langle1|$ and $|1\rangle_a$ generates the Fock spaces, which will be denoted as $\cF_a=\bigoplus^\infty_{L=0}(\cF_a)_L$ and $\bcF_a=\bigoplus^\infty_{L=0}(\bcF_a)_L$ respectively. The gradation of the Fock spaces is defined naturally: $\deg d^\pm_k=k$ for $\cF_a$ and $\deg d^\pm_k=-k$ for $\bcF_a$.

The vacuum vectors define the normal ordering symbol~$\lcolon\cdots\rcolon$. We assume the it will put all elements $d^\pm_k$ with $k>0$ to the right of those with $k<0$. On the other hand, it will be convenient to assume that it does not affect the order of the zero mode operators $\hat a$ and~$\d_a$, so that $\lcolon AB\rcolon$ may not coincide with $\lcolon BA\rcolon$ if the operators contain the zero modes.

Define the vertex operators
\eq$$
\lambda_\pm(x)=\exp\sum_{k\ne0}{d^\pm_kz^{-k}\over k}.
\label{lambdapm-def}
$$
Their operator products read
\eq$$
\Aligned{
\lambda_\pm(z_1)\lambda_\pm(z_2)
&=\lcolon\lambda_\pm(z_1)\lambda_\pm(z_2)\rcolon,
\\
\lambda_+(z_1)\lambda_-(z_2)
&=\lambda_-(z_2)\lambda_+(z_1)=f\left(z_2\over z_1\right)\lcolon\lambda_+(z_1)\lambda_-(z_2)\rcolon
\quad(z_2\ne\pm z_1).
}\label{lambda-norm}
$$
Here
\eq$$
f(z)={(z+q)(z-q^{-1})\over(z^2-1)},
\qquad
q=\e^{-\i\pi r}.
\label{f(z)-def}
$$
The currents
\eq$$
t(z)=\e^{\i\pi\hat a}\lambda_-(z)+\e^{-\i\pi\hat a}\lambda_+(z),
\qquad
s(z)=\lcolon\lambda_-(z)\lambda_+(-z)\rcolon.
\label{ts-def}
$$
generate the algebra ${\it SVir}_{q,-q}$ described in detail in~\cite{Lashkevich:2013yja}.

The matrix elements
\eq$$
J_a(X)={}_a\langle1|t(X)|1\rangle_a,
\qquad
t(X)=t(x_1)t(x_2)\cdots t(x_N),
\label{Ja-ff}
$$
define a local operator $V_a(x)$, which is nothing but an exponential operator divided by its vacuum expectation value:
\eq$$
\e^{\i\alpha\varphi(x)}=G_\alpha V_a(x),
\qquad
a={1\over2}-{\alpha\over\beta^{-1}-\beta}.
\label{Va-def}
$$
where $G_\alpha=\langle\e^{\i\alpha\varphi(x)}\rangle$ is the vacuum expectation value of the exponential operator~\cite{Lukyanov:1996jj}.

Equation (\ref{Ja-ff}) admits a simple generalization. Consider a commutative algebra $\cA$ generated by the elements $c_{-1},c_{-2},\ldots$. This algebra admits a natural grading $\cA=\bigoplus^\infty_{L=0}\cA_L$ by assuming $\deg c_{-k}=k$. We will need two representations of the algebra $\cA$ in terms of the free bosons $d^\pm_k$:
\eq$$
\pi(c_{-k})={d^-_k-d^+_k\over A^+_k},
\qquad
\bar\pi(c_{-k})={d^-_{-k}-d^+_{-k}\over A^+_k}.
\label{pibpi-def}
$$
The vectors
\eq$$
{}_a\langle h|={}_a\langle1|\pi(h),
\qquad
|h\rangle_a=\bar\pi(h)|1\rangle_a,
\qquad
h\in\cA,
\label{hstates-def}
$$
span subspaces~$\cF^\cA_a$ and~$\bcF^\cA_a$ in the spaces $\cF_a$ and $\bcF_a$ respectively, which we call $\cA$\=/subspaces. The grading of $\cA$ and the gradings of the Fock spaces are consistent.

For any pair of elements $h,h'\in\cA$ we define a set of functions
\eq$$
J^{h\bh'}_a(X)={}_a\langle h|t(X)|h'\rangle_a.
\label{Jhh'a-def}
$$
By substituting these functions into~(\ref{FJ-rel}) we define a set of functions $F^{h\bar h'}_a(\theta_1,\ldots,\theta_N)$, which turn out to be form factors of an operator. We will denote this operator~$V^{h\bar h'}_a(x)$. In~\cite{Feigin:2008hs} it was argued that for generic values of~$a$, if $h\in\cA_L$, $h'\in\cA_{\bar L}$, the operator $V^{h\bar h'}_a(x)$ is a linear combination of the level $(L-k,\bar L-k)$ descendants of the operator $V_a(x)$ with $0\le k\le\min(L,\bar L)$ with a nonzero highest level component.

The matrix elements on the r.h.s.\ of~(\ref{Jhh'a-def}) are calculated by means of the following commutation relation:
\subeq{\label{pibpi-props}
\Align$$
[\pi(c_{-k}),\lambda_\pm(z)]
&=(\mp)^{k+1}z^k\lambda_\pm(z),
\label{pilambda-commut}
\\
[\bar\pi(c_{-k}),\lambda_\pm(z)]
&=-(\pm)^{k+1}z^{-k}\lambda_\pm(z),
\label{bpilambda-commut}
\\
[\pi(c_{-k}),\bar\pi(c_{-l})]
&=-(1+(-1)^k)k(A^+_k)^{-1}\delta_{kl},
\label{pibpi-commut}
$$}
and the evident identities
\eq$$
\pi(c_{-k})|1\rangle_a=0,
\qquad
{}_a\langle1|\bar\pi(c_{-k})=0.
\label{pibpi-zero}
$$
Eqs.\ (\ref{pibpi-props}), (\ref{pibpi-zero}) result in simple rules to obtain $J$\=/functions explicitly. These rules are listed in Appendix~\ref{Jfunc-appendix}.

Note that, as it was explained in~\cite{Feigin:2008hs}, every element $c_{1-2k}$ with $k$ positive integer corresponds to the spin $\sigma=\pm(2k-1)$ integrals of motion $I_{\pm(2k-1)}$. That is,
$$
J^{c_{1-2k}h\bh'}_a(X)=J^{h\bh'}(X)\sum^N_{i=1}x_i^{2k-1},
\quad
J^{h\overline{c_{1-2k}h'}}_a(X)=J^{h\bh'}(X)\sum^N_{i=1}x_i^{1-2k},
$$
and, since $\sum\e^{\sigma\theta_i}$ is an eigenvalue of~$I_\sigma$, we have
$$
V^{c_{1-2k}h\bh'}_a(x)=\const\cdot[V^{h\bh'}_a(x),I_{2k-1}],
\quad
V^{h\,\overline{c_{1-2k}h'}}_a(x)=\const\cdot[V^{h,\bh'}_a(x),I_{1-2k}].
$$
On the contrary, the elements $c_{-2k}$ act nontrivially and produce essentially new physical operators. The property (\ref{pibpi-commut}), which mixes the two physical chiralities, seems to be especially important, since it is responsible for most of the properties of the screening operators described in the next section.

Dealing with $J$\=/functions we have to constantly keep in mind two main properties. First, they are quasiperiodic in~$a$:
\eq$$
J^{h\bar h'}_{a+1}(X)=(-1)^NJ^{h\bar h'}_a(X).
\label{Jhh'a-quasiperidisity}
$$
Second, they satisfy a kind of reflection relation. In~\cite{Feigin:2008hs} it was proved that there exists such a continuous family of linear maps $r_a:\cA\to\cA$ consistent with the grading that
\eq$$
J^{h\bh'}_a(X)=J^{r_a(h)\,\overline{r_{-a}(h')}}_{-a}(X).
\label{Jhh'a-reflection}
$$
For the unit element of $\cA$ this map is trivial, $r_a(1)=1$, so that the operator $V_a(x)$ satisfies the relations
\eq$$
V_a(x)=V_{-a}(x)=V_{a+2}(x),
\label{Va-equations}
$$

Now consider the case of higher breathers. Let $F^{(\nu)}_\cO(\vartheta,\theta_1,\ldots,\theta_N)=F_\cO(\vartheta,\theta_1,\ldots,\theta_N)_{\nu1\ldots1}$ be the form factor of the operator $\cO(x)$ with one $\nu$th breather with the rapidity $\vartheta$ and $N$ first breathers with the rapidities $\theta_1,\ldots,\theta_N$. We have the fusion relation~\cite{Smirnov:1992vz}
\eq$$
F^{(\nu+1)}_\cO(\vartheta,\theta_1,\ldots,\theta_N)
=\left(\i\Gamma^{(\nu)}\right)^{-1}\Res_{\vartheta'=\vartheta}
F_\cO^{(\nu)}\left(\vartheta+{\textstyle{\i\pi(r-1)\over2}},\vartheta'-{\textstyle{\i\pi(r-1)\nu\over2}},\theta_1,\ldots,\theta_N\right),
\label{F-fusion}
$$
where the coefficient $\Gamma^{(\nu)}$ was defined in~(\ref{Gamma(nu)-def}).

The decomposition analogous to (\ref{FJ-rel}) for these form factors reads
\eq$$
F^{(\nu)}_\cO(\vartheta,\theta_1,\ldots,\theta_N)
=\rho^N\rho^{(\nu)}J^{(\nu)}_\cO\left(\e^\vartheta;\e^{\theta_1},\ldots,\e^{\theta_N}\right)
\prod^N_{i=1}R^{(\nu)}(\vartheta-\theta_i)\prod^N_{i<j}R(\theta_i-\theta_j),
\label{FJ(nu)-rel}
$$
where the exact form of $\rho^{(\nu)}$ and $R^{(\nu)}(\theta)$ can be found in Appendix~\ref{reference-appendix} as well. The relation (\ref{F-fusion}) can be rewritten in terms of the $J$\=/functions as follows:
\eq$$
J^{(\nu+1)}_\cO(z;X)=J^{(\nu)}_\cO(z\tq^{1/2};z\tq^{-\nu/2},X).
\label{J-fusion}
$$
Here
\eq$$
\tq=\e^{-\i\pi}q^{-1}=\e^{-\i\pi(1-r)}.
\label{tq-def}
$$
Now formulate the fusion relation in terms of the free field realization. Define the currents $t^{(\nu)}(z)$ according to
\eq$$
t^{(1)}(z)=t(z),
\qquad
t^{(\nu+1)}(z)=t^{(\nu)}(z\tq^{1/2})t(z\tq^{-\nu/2}).
\label{t-fusion}
$$
Then, evidently,
\eq$$
J^{(\nu)h\bh'}_a(z;X)={}_a\langle h|t^{(\nu)}(z)t(X)|h'\rangle_a.
\label{Jt(nu)-rel}
$$
Explicitly, the currents $t^{(\nu)}(z)$ read
\eq$$
t^{(\nu)}(z)=\prod^{\nu-1}_{i=0}t(\tq^{{\nu-1\over2}-i}z)
=\sum^\nu_{j=0}\e^{\i\pi(\nu-2j)\hat a}f^{(\nu)}_j\lambda^{(\nu)}_j(z),
\label{tnu-def}
$$
where
\Gather$$
\lambda^{(\nu)}_j(z)
=\lcolon\prod^{\nu-1}_{i=j}\lambda_-(\tq^{-(\nu-1)/2+i}z)\prod^{j-1}_{i=0}\lambda_+(\tq^{-(\nu-1)/2+i}z)\rcolon,
\label{lambdanu-def}
\\
f^{(\nu)}_j=\prod^{j-1}_{i=0}\prod^{\nu-1}_{i'=j}f(\tq^{i'-i})
={\prod^{j-1}_{i=0}(1+\tq^i)\prod^\nu_{i=\nu-j+1}(1-\tq^i)
\over\prod^{\nu-1}_{i=\nu-j}(1+\tq^i)\prod^j_{i=1}(1-\tq^i)}.
\label{fnu-def}
$$
Note that
\eq$$
f^{(\nu)}_j=f^{(\nu)}_{\nu-j}.
\label{fnu-sym}
$$
In the simplest case $\nu=1$ we have $\lambda^{(1)}_0(z)=\lambda_-(z)$, $\lambda^{(1)}_1(z)=\lambda_+(z)$.

Now let us formulate the condition of compatibility with the reduction. Let $r$ be one of the values~(\ref{rs-values}). An operator $\cO(x)$ is compatible with the reduction if and only if
\eq$$
F^{(2s-2)}_\cO(\vartheta,\theta_1,\ldots,\theta_N)=F^{(1)}_\cO(\vartheta,\theta_1,\ldots,\theta_N)
\quad\forall N,\vartheta,\theta_i.
\label{redcompat}
$$
In the $J$\=/function language the condition reads
\eq$$
C_s\prod^N_{i=1}h\left(z\over x_i\right)J^{(2s-2)}_\cO(z,x_1,\ldots,x_N)=J_\cO(z,x_1,\ldots,x_N)
\quad\forall N,z,x_i.
\label{redcompat-J}
$$
Here
\eq$$
C_s={(-1)^s\over2s-1}\tg{\pi\over2s-1},
\qquad
h(z)={(z-\tq)(z-\tq^{-1})\over(z+1)^2}.
\label{Ch-def}
$$
Identities necessary to derive it can be found in Appendix~\ref{reference-appendix}.

The condition (\ref{redcompat}) or (\ref{redcompat-J}) has the form of infinite number of equations. Substituting the Ansatz (\ref{Jhh'a-def}), (\ref{Jt(nu)-rel}) for the $J$\=/functions, we get a system of equations that poorly can be solved for general values of $h,h'$. We need to simplify the condition. In what follows we challenge this problem. We will see that just the algebraic structure, which at the beginning seemed to be nothing but a complicated way to write down rather simple functions, makes it possible to reduce the problem to an $N$\=/independent set of equations. Below we construct a wide class of solutions and, among them, a set of rather useful and naturally defined local operators.

\section{The screening algebra}
\label{scralg-sec}

In this section we define an algebra which we call the screening algebra. It extends the algebra introduced in~\cite{Lashkevich:2014rua}, which is generated by the screening currents. Here we introduce what can be called `antiscreening' currents. In this section we produce the relations of the screening algebra and the relations between modes of screening currents and the $t^{(\nu)}(z)$ currents. An explanation, how the relations produced in this section are proved, is placed in Appendix~\ref{screening-rel-appendix}. Later we will use the antiscreening currents to generate $\cA$\=/vectors with some particular property necessary for reduction.

Below we will consider currents, which depend on a complex parameter $z$, say $A(z)$. These currents are all constructed of the combinations $d^\pm_kz^{-k}$ and two shift operators
\eq$$
\delta=\e^{(1-r)\d_a}\>:\bcF_a\to\bcF_{a-(1-r)},
\qquad
\tdelta=\e^{r\d_a}\>:\bcF_a\to\bcF_{a-r}.
\label{delta-def}
$$
Besides, we will need the modes or Laurent components $A_k$ of the currents:
\eq$$
A(z)=\sum_{k\in\Z}A_kz^{-k},
\qquad
A_k=\oint{dz\over2\pi\i}\,z^{k-1}A(z).
\label{Sk-def}
$$
As usual, we always assume that in the product $A_kB_l$ the contour of the first operator envelops all the poles of the operator product $A(z)B(z')$, while the contour of the second one leaves them outside.

The screening currents are defined as
\Align$$
S(z)
&=\delta\lcolon\exp\sum_{k\ne0}{d^-_k-d^+_k\over k(q^{k/2}-q^{-k/2})}z^{-k}\rcolon,
\label{S-def}
\\
\tS(z)
&=\tdelta\lcolon\exp\sum_{k\ne0}{d^-_k-d^+_k\over k(\tq^{k/2}-\tq^{-k/2})}z^{-k}\rcolon,
\label{tS-def}
$$
The corresponding antiscreening currents are
\Align$$
S^+(z)
&=\lcolon S^{-1}(z)\rcolon=\delta^{-1}\lcolon\exp\sum_{k\ne0}{d^+_k-d^-_k\over k(q^{k/2}-q^{-k/2})}z^{-k}\rcolon,
\label{S+-def}
\\
\tS^+(z)
&=\lcolon\tS^{-1}(z)\rcolon
=\tdelta^{-1}\lcolon\exp\sum_{k\ne0}{d^+_k-d^-_k\over k(\tq^{k/2}-\tq^{-k/2})}z^{-k}\rcolon.
\label{tsigma-def}
$$
At last, we need some extra currents, which will appear in the commutation relations of the screening currents:
\Align$$
\eta(z)
&=\exp\sum_{k\in2\Z+1}{2(d^+_k-d^-_k)z^{-k}\over k(q^{k/2}-q^{-k/2})},
\label{eta-def}
\\
\teta(z)
&=\exp\sum_{k\in2\Z+1}{2(d^+_k-d^-_k)z^{-k}\over k(\tq^{k/2}-\tq^{-k/2})},
\label{teta-def}
\\
\epsilon(z)
&=\delta\tdelta\exp\sum_{k\in2\Z+1}{2(d^-_k-d^+_k)\over k(q^k-q^{-k})}z^{-k}.
\label{epsilon-def}
\\
\epsilon^+(z)
&=\epsilon^{-1}(z)=\delta^{-2}\tdelta^{-2}\epsilon(-z).
\label{epsilon+-def}
$$
Note that these extra currents are expressed in terms of the differences $d^-_k-d^+_k$ with $k$ odd. This results in the fact that their modes $\eta_k$, $\teta_k$, $\epsilon_k$ are central in the algebra, i.e.\ commute with all elements of the screening algebra.

The modes of the defined currents act on Fock spaces as follows:
\eq$$
\Aligned{
S_k:\
&(\bcF_a)_L\to(\bcF_{a-(1-r)})_{L-k},
\qquad
&S^+_k:\
&(\bcF_a)_L\to(\bcF_{a+(1-r)})_{L-k},
\\
\tS_k:\
&(\bcF_a)_L\to(\bcF_{a-r})_{L-k},
&\tS^+_k:\
&(\bcF_a)_L\to(\bcF_{a+r})_{L-k},
\\
\epsilon_k:\
&(\bcF_a)_L\to(\bcF_{a-1})_{L-k},
&\epsilon^+_k:\
&(\bcF_a)_L\to(\bcF_{a+1})_{L-k},
\\
\span\span
\eta_k,\teta_k:\
(\bcF_a)_L\to(\bcF_a)_{L-k}.
}\label{FF-action}
$$

The relations of the screening algebra look like:
\subeq{\label{screening-commut}
\Align$$
S_kS_l+S_{l+2}S_{k-2}
&=0,
\qquad
S^+_kS^+_l+S^+_{l+2}S^+_{k-2}=0,
\label{SS-commut}
\\
\tS_k\tS_l+\tS_{l+2}\tS_{k-2}
&=0,
\qquad
\tS^+_k\tS^+_l+\tS^+_{l+2}\tS^+_{k-2}=0,
\label{tStS-commut}
\\
S^+_kS_l+S_{l-2}S^+_{k+2}
&={1\over2}\left(\delta_{k+l,0}+(-)^l\eta_{k+l}\right),
\label{S+S-commut}
\\
\tS^+_k\tS_l+\tS_{l-2}\tS^+_{k+2}
&={1\over2}\left(\delta_{k+l,0}+(-)^l\teta_{k+l}\right),
\label{tS+tS-commut}
\\
S_k\tS^+_l-\tS^+_{l+2}S_{k-2}
&=0,
\qquad
\tS_kS^+_l-S^+_{l+2}\tS_{k-2}=0,
\label{StS+-commut}
\\
S_k\tS_l-\tS_{l-2}S_{k+2}
&={\i^l\over2}\left(q^{(k+l)/2}+(-1)^lq^{-(k+l)/2}\right)\epsilon_{k+l}.
\label{StS-commut}
\\
S^+_k\tS^+_l-\tS^+_{l-2}S^+_{k+2}
&={\i^l\over2}\left(q^{(k+l)/2}+(-1)^lq^{-(k+l)/2}\right)\epsilon^+_{k+l}.
\label{S+tS+-commut}
$$}

An important feature of the screening operators' modes $S_k$, $\tS_k$ is the fact that they have nice commutation relations with the particle\-/creating currents~$t^{(\nu)}(z)$. In fact, we will restrict ourselves to the commutation relation of~$S_k$. That of the modes $\tS_k$ can be obtained straightforwardly by the substitution $r\to1-r$, but will not be used below. The algebraic part of our construction is symmetric with respect to this substitution. The explicit asymmetry of the results of this paper with respect to this substitution is related to the addressed problem: the reduction to the perturbed minimal $M(2,2s+1)$ models, corresponding to special values of~$r$.

The following currents will be necessary:
\eq$$
\sigma^{(\nu)}_j(z)=\lcolon\lambda^{(\nu)}_j(z)S(-\i\tq^{j-\nu/2}z)\rcolon
=\lcolon\lambda^{(\nu)}_{j+1}(z)S(\i\tq^{j+1-\nu/2}z)\rcolon,
\qquad
j=0,1,\ldots,\nu-1.
\label{sigma-def}
$$
These currents generalize the current $\sigma(z)=\sigma^{(1)}_0(z)$ introduced in~\cite{Lashkevich:2013yja}. We have
\eq$$\relax
[S_k,t^{(\nu)}(z)]=z^k\sum^{\nu-1}_{j=0}\beta^{(\nu)}_j\sigma^{(\nu)}_j(z)\,
\e^{\i\pi(\nu-2j-1)\left(\hat a+{1-r\over2}(k-1)\right)}\cos\left(\pi\hat a-{\pi r\over2}(k-1)\right).
\label{St-commut}
$$
Here
\eq$$
\beta^{(\nu)}_j=\i f^{(\nu)}_j\left(\tq^{\nu-j\over2}-\tq^{j-\nu\over2}\right)\left(\tq^{j\over2}+\tq^{-{j\over2}}\right),
\qquad
\beta^{(\nu)}_{\nu-1-j}=\beta^{(\nu)}_j.
\label{varphi(nu)j-def}
$$

We will be interested in the particular values of $a$, for which there exist special operators, which commute with the currents~$t^{(\nu)}(z)$, the so called screening operators. Let
\eq$$
a_{mn}={rm\over2}+{(1-r)n\over2}.
\label{amn-def}
$$
We will use the notation $\cF_{mn}=\cF_{a_{mn}}$, $\bcF_{mn}=\bcF_{a_{mn}}$, $V^{h\bh'}_{mn}(x)=V^{h\bh'}_{a_{mn}}(x)$ etc.\ by omitting the letter $a$ if $a=a_{mn}$. In~\cite{Lashkevich:2013yja} an infinite set of screening operators $Q^{(s)}$, $\tQ^{(s)}$ ($s=1,2,\ldots$) was defined. Here we need only two simplest ones, $Q^{(1)}=\Sigma$ and $\tQ^{(1)}=\tSigma$:
\eq$$
\Sigma|_{\bcF_{mn}}=S_{m-n+1},
\qquad
\tSigma|_{\bcF_{mn}}=\tS_{n-m+1}
\label{Sigma-def}
$$
The screening operators act on the Fock spaces as
\eq$$
\Sigma:(\bcF_{mn})_l\to(\bcF_{m,n-2})_{l-m+n-1},
\qquad
\tSigma:(\bcF_{mn})_l\to(\bcF_{m-2,n})_{l+m-n-1}.
\label{Sigma-action}
$$
They satisfy the relations
\eq$$
\Sigma^2=0,
\qquad
\tSigma^2=0,
\qquad
[\tSigma,\Sigma]={\i^{m-n+1}\over2}(1-(-1)^{m-n})\epsilon_0.
\label{Sigma-rel}
$$
The main property of the screening operators is that they commute with the currents~$t^{(\nu)}(z)$:
\eq$$
\Aligned{
[\Sigma,t^{(\nu)}(z)]|_{\bcF_{mn}}
&=0,\quad\text{if $n\in2\Z+1$,}
\\
[\tSigma,t^{(\nu)}(z)]|_{\bcF_{mn}}
&=0,\quad\text{if $m\in2\Z+1$.}
}\label{Sigma-t-commut}
$$

Since all the operators discussed in this section are expressed in terms of the differences $d^-_k-d^+_k$, they map $\cA$\=/vectors into $\cA$\=/vectors and, in particular, make it possible to create $\cA$\=/states from the vacuum. We may define the vectors
\eq$$
\Aligned{
{}_a\langle\ft_{-k_1,\ldots,-k_M}|
&={}_{a-rM}\langle1|S_{k_1}\cdots S_{k_M},
&
{}_a\langle\tilde\ft_{-k_1,\ldots,-k_M}|
&={}_{a-(1-r)M}\langle1|\tS_{k_1}\cdots\tS_{k_M},
\\
{}_a\langle\ft^+_{-k_1,\ldots,-k_M}|
&={}_{a+rM}\langle1|S^+_{k_1}\cdots S^+_{k_M},
&{}_a\langle\tilde\ft^+_{-k_1,\ldots,-k_M}|
&={}_{a+(1-r)M}\langle1|\tS^+_{k_1}\cdots\tS^+_{k_M},
\\
\span\span\span
{}_a\langle\eta_{-k}|
={}_a\langle1|\eta_k,
\quad
{}_a\langle\teta_{-k}|
={}_a\langle1|\teta_k,
\quad
{}_a\langle\epsilon_{-k}|
={}_{a-1}\langle1|\epsilon_k
}\label{scr-vectors}
$$
Here everywhere $k\ge0$, $k_1\ge0$, $k_{i+1}\ge k_i-1$, $\sum^M_{i=1}k_i\ge0$. These vectors define elements $\ft_{-k_1,\ldots,-k_M},\allowbreak\ldots,\allowbreak\teta_{-k},\allowbreak\epsilon_{-k}\in\cA$. More explicitly, these elements are defined by means of series~(\ref{screl-series}). Here we give the formulas in the case $M=1$, since we need them later:
\eq$$
\sum^\infty_{k=0}\ft_{-k}z^k
=\exp\sum^\infty_{k=1}B_kc_{-k}z^k,
\quad
\sum^\infty_{k=0}\ft^+_{-k}z^k
=\exp\sum^\infty_{k=1}(-B_k)c_{-k}z^k,
\label{ftk-def}
$$
where
\eq$$
B_k={q^{k/2}-(-1)^kq^{-k/2}\over k}.
\label{Bk-def}
$$
The elements $\tft_k$ and $\tft^+_k$ are defined by the same equalities, where the coefficients $B_k$ are substituted by $\tB_k=k^{-1}(\tq^{k/2}-(-1)^k\tq^{-k/2})$.

\section{The \texorpdfstring{$\tau^{(\nu)}(z)$}{T(nu)(z)} currents}
\label{tau-sec}

Consider the currents
\eq$$
\tau^{(\nu)}(z)
=\tdelta^{-1}\lcolon\exp\sum_{k\ne0}{\tq^{\nu k/2}d^+_k-\tq^{-\nu k/2}d^-_k\over k(\tq^{k/2}-\tq^{-k/2})}z^{-k}\rcolon,
\qquad
\nu=1,2,3,\ldots
\label{tau-def}
$$
These currents are related with the fusion $t$\=/currents by the identity:
\eq$$
\tau^{(\nu)}(z)\tS_n+(-1)^\nu z^2\tS_{n-2}\tau^{(\nu)}(z)=K^{(\nu)}z^nt^{(\nu)}(z)|_{\bcF_{1n}},
\label{tauS-commut}
$$
where
\eq$$
K^{(\nu)}={\i^\nu\tq^{\nu/2}\over1-\tq^\nu}\prod^{\nu-1}_{j=1}{1+\tq^j\over1-\tq^j}
={1\over2\sin{\pi(1-r)\nu\over2}}\prod^{\nu-1}_{j=1}\ctg{\pi(1-r)j\over2}.
\label{Knu-def}
$$
In particular, $K^{(1)}=B_1^{-1}$. Note that $\tS_n=\tSigma|_{\bcF_{1n}}$ and $\tS_{n-2}=\tSigma|_{\bcF_{3n}}$, so that the identity~(\ref{tauS-commut}) can be rewritten as
\eq$$
\tau^{(\nu)}(z)\tSigma+(-1)^\nu z^2\tSigma\tau^{(\nu)}(z)|_{\bcF_{1n}}
=K^{(\nu)}z^nt^{(\nu)}(z)|_{\bcF_{1n}},
\tag{\ref{tauS-commut}$'$}
$$
or, more generally, as
\eq$$
\tau^{(\nu)}(z)\tSigma+(-1)^\nu z^2\tSigma\tau^{(\nu)}(z)|_{\bcF_{mn}}
=(-1)^{m-1\over2}K^{(\nu)}z^{n-m+1}t^{(\nu)}(z)|_{\bcF_{mn}},
\quad\text{if $m\in2\Z+1$.}
\tag{\ref{tauS-commut}$''$}
$$

{\bf Proof of identity (\ref{tauS-commut}).} It is easy to check that
\eq$$
\tau^{(\nu)}(z)\tS(z')=(-1)^{\nu+1}{z^2\over z^{\prime2}}\tS(z')\tau^{(\nu)}(z)
=g^{(\nu)}\left(z'\over z\right)\lcolon\tau^{(\nu)}(z)\tS(z')\rcolon,
\label{tautS-prod}
$$
where
\eq$$
g^{(\nu)}(z)={\prod^{\nu-1}_{j=1}(1+\tq^{\nu/2-j}z)\over\prod^\nu_{j=0}(1-\tq^{\nu/2-j}z)}.
\label{gnu-def}
$$
The l.h.s.\ of (\ref{tauS-commut}) is nothing but an integral of the l.h.s.\ of~(\ref{tautS-prod}):
$$
\tau^{(\nu)}(z)\tS_n+(-1)^\nu z^2\tS_{n-2}\tau^{(\nu)}(z)
=-\oint_{\cC^{(\nu)}_z}{dz'\over2\pi\i}\,z^{\prime n-1}\tau^{(\nu)}(z)\tS(z').
$$
The contour $\cC^{(\nu)}_z$ encircles the poles $z\tq^{j-\nu/2}$ ($j=0,1,\ldots,\nu$) of the integrand. By substituting (\ref{tautS-prod}) into the r.h.s.\ we obtain
\eq$$
-\sum^\nu_{j=0}z(z\tq^{j-\nu/2})^{n-1}r^{(\nu)}_j\lcolon\tau^{(\nu)}(z)\tS(z\tq^{j-\nu/2})\rcolon.
\label{rightsum}
$$
Here
\eq$$
r^{(\nu)}_j=\Res_{z=\tq^{j-\nu/2}}g^{(\nu)}(z)=\i^\nu(-1)^{j+1}K^{(\nu)}f^{(\nu)}_j.
\label{rnu-def}
$$
The last equality is proved by a straightforward calculation. Besides, it is easy to check that
$$
\lcolon\tau^{(\nu)}(z)\tS(z\tq^{j-\nu/2})\rcolon=\lambda^{(\nu)}_j(z).
$$
At last, $\i^\nu(-)^j\tq^{j-\nu/2}=\e^{\i\pi(\nu-2j)a_{1n}}$. This reduces~(\ref{rightsum}) to the r.h.s.\ of~(\ref{tauS-commut}).~\vrule width 5pt

Note the following properties of the currents~$\tau^{(\nu)}(z)$:
\Align$$
\tau^{(\nu)}(\tq^{\mp1/2}z)t(\tq^{\pm\nu/2}z)
&={1-\tq^{\pm(\nu+1)}\over1+\tq^{\pm\nu}}\tau^{(\nu+1)}(z)\e^{\pm\i\pi a},
\label{taut-fusion}
\\
t(z')\tau^{(\nu)}(z)
&=\tau^{(\nu)}(z)t(z').
\label{taut-commut}
$$
They immediately follow from the relations
\Gather*$$
\lambda_\pm(z')\tau^{(\nu)}(z)
=-\tq^{\pm1}\tau^{(\nu)}(z)\lambda_\pm(z')
={z'-\tq^{\pm(\nu+1)/2}z\over z'+\tq^{\pm(\nu-1)/2}z}
\lcolon\lambda_\pm(z')\tau^{(\nu)}(z)\rcolon
\\
\lcolon\tau^{(\nu)}(\tq^{\mp1/2}z)\lambda_\mp(\tq^{\pm\nu/2}z)\rcolon
=\tau^{(\nu+1)}(z).
$$
Notice that equation (\ref{taut-fusion}) provides an alternative (inductive) proof of~(\ref{tauS-commut}). The base of induction is given by eq.~(\ref{tauS-commut}) at $\nu=1$ proved directly.

There is a simple example of an application of eq.~(\ref{tauS-commut}). It makes it possible to easily prove that all (except zero-particle) form factors of the operator $V_{11}(x)$ vanish and, hence, it is the unit operator. Indeed, for $n=1$ we have
\Multline*$$
{}_{11}\langle1|t(z)t(X)|1\rangle_{11}
=K^{(1)}\left({}_{11}\langle1|\tau^{(1)}(z)\tS_1t(X)|1\rangle_{11}
-{}_{11}\langle1|z^2\tS_{-1}\tau^{(1)}(z)t(X)|1\rangle_{11}\right)
\\
=K^{(1)}\left({}_{11}\langle1|\tau^{(1)}(z)t(X)\tS_1|1\rangle_{11}
-{}_{11}\langle1|z^2\tS_{-1}\tau^{(1)}(z)t(X)|1\rangle_{11}\right)=0,
$$
since ${}_a\langle1|\tS_{-1}=0$ and $\tS_1|1\rangle_a=0$.

The currents $\tau^{(\nu)}(z)$ successfully substitute the currents $\sigma^{(\nu)}_j(z)$ introduced in the previous section. It is easy to check that
\eq$$
\sigma^{(\nu)}_j(z)=\lcolon\tau^{(\nu)}(z)\epsilon\left(\tq^{j-{\nu-1\over2}}z\right)\rcolon.
\label{sigma-tau-rel}
$$
By using the operator products
\eq$$
\epsilon(z')\tau^{(\nu)}(z)=(-1)^\nu\tau^{(\nu)}(z)\epsilon(z')
=\lcolon\epsilon(z')\tau^{(\nu)}(z)\rcolon
\prod^{\nu-1}_{j=0}{z'+\tq^{\left({\nu-1\over2}-j\right)k}z\over z'-\tq^{\left({\nu-1\over2}-j\right)k}z}
\label{epsilon-tau-OP}
$$
eq.~(\ref{sigma-tau-rel}) can be rewritten as
\eq$$
\epsilon_0\tau^{(\nu)}(z)-(-1)^\nu\tau^{(\nu)}(z)\epsilon_0
=\i^{1-\nu}K^{(\nu)}\sum^{\nu-1}_{j=0}(-1)^j\beta^{(\nu)}_j\sigma^{(\nu)}_j(z).
\label{sigma-tau-epsilon0}
$$
The r.h.s.\ resembles the r.h.s.\ of eq.~(\ref{St-commut}). Let us take the $\hat a$\=/derivative of the latter and then specialize to $a=a_{mn}$ with arbitrary odd values of~$m,n$. We obtain
\eq$$
[\Sigma,t^{(\nu)\prime}(z)]|_{\bcF_{mn}}
=\i^{(\nu-1)m+n}\pi z^{m-n+1}\sum^{\nu-1}_{j=0}(-1)^j\beta^{(\nu)}_j\sigma^{(\nu)}_j(z)
\quad\text{for $m,n\in2\Z+1$}.
\label{S't-mnodd-commut}
$$
Here the prime at $t^{(\nu)}(z)$ means the $\hat a$\=/derivative. Comparison of (\ref{sigma-tau-epsilon0}) and (\ref{S't-mnodd-commut}) gives the identity
\Multline$$
[\Sigma,t^{(\nu)\prime}(z)]|_{\bcF_{mn}}
=(-1)^{n+1+(m+1)(\nu-1)\over2}{\pi\over K^{(\nu)}}z^{m-n+1}(\epsilon_0\tau^{(\nu)}(z)-(-1)^\nu\tau^{(\nu)}(z)\epsilon_0)
\\
\text{for $m,n\in2\Z+1$.}
\label{S't-mnodd-tau}
$$
This equation will be used later to derive eq.~(\ref{V'hh'-delta}).

\section{Reduction compatibility condition}
\label{compatcond-sec}

Now let us specialize to the case~(\ref{rs-values}), which corresponds to the $\Phi_{13}$\=/perturbations of the $M(2,2s+1)$ minimal conformal models. In this case
\eq$$
\tq=\omega\equiv\e^{2\pi\i/(2s-1)}.
\label{tqs-values}
$$
The limit $\tq\to\omega$ is singular in the free field representation. Modes at particular values of $k$ are undefined. It is convenient to have special notation for sets of these `dangerous' values of the mode label~$k$:
\eq$$
\bbD_s=(2s-1)\Z,
\qquad
\bbDodd_s=(2s-1)(2\Z-1),
\qquad
\bbDeven_s=(2s-1)\cdot2\Z.
\label{badsets-def}
$$
We will also add subscripts `$\ne0$', `$>0$' to narrow the sets to nonzero or positive numbers.

For $k\not\in\bbD_s$ the elements $d^\pm_k$ are regular in the limit $\tq\to\omega$, while for $k\in\bbD_s$ they become commuting. It is worse that the elements $\pi(c_{-k})$ and $\bar\pi(c_{-k})$ get infinite factors $(A^+_k)^{-1}$ for $k\in\bbD_s$. To define the algebra correctly, it is convenient to use the elements
\eq$$
\bfd^\pm_k={d^\pm_k\over\tq^{k/2}-\tq^{-k/2}},
\label{bfd-def}
$$
In the limit $\tq\to\omega$ the operators $\bfd^\pm_k$ are undefined for $k\in\bbDodd_s$, but we may consider the operators $\bfD_k=\bfd^-_k-\bfd^+_k$, which are well-defined. Moreover, they allow one to define the representatives of~$c_{-k}$:
\eq$$
\pi(c_{-k})=-{\bfD_k\over2},
\qquad
\bar\pi(c_{-k})=-{\bfD_{-k}\over2},
\qquad
k\in\bbDodd_s.
\label{ck-Dodd}
$$
For $k\in\bbDeven_s$, though $\bfd^\pm_k$ are well-defined, the operators $\pi(c_{-k})$ and $\bar\pi(c_{-k})$ are completely undefined and the corresponding elements $c_{-k}$ must be excluded from the spectrum. Formally, we may consider the elements $\check c_{-k}=(\tq^{k/2}-(-1)^k\tq^{-k/2})c_{-k}$ Then $\pi(\check c_{-k})=\bfd^-_k-\bfd^+_k$, $\bar\pi(\check c_{-k})=\bfd^-_{-k}-\bfd^+_{-k}$ for all values of~$k$. These elements are well\-/defined in the limit $\tq\to\omega$. Thus, the algebra~$\check\cA$ generated by the elements~$\check c_{-k}$ coincides with $\cA$ for generic values of~$\tq$ and remains sensible in the limit $\tq\to\omega$. The elements $\check c_{-k}$ for $k\in\bbDeven_s$ produce no new physical operators since
$$
[\pi(\check c_{-k}),t(z)]=[\pi(\check c_{-k}),t(z)]=0
\quad\text{for $k\in\bbDeven_s$.}
$$

More formally, the Heisenberg algebra in the limit $\tq\to\omega$ consists of the elements $\d_a$, $\hat a$, $d^\pm_k$ ($k\in\Z\setminus\bbD_s$), $d^+_k$, $\bfD_k$ ($k\in\bbDodd_s$), $\bfd^\pm_k$ ($k\in\bbDeven_{s,\ne0}$) with the commutation relations~(\ref{dpm-commut}) for $d^\pm_k$ and the following nonvanishing commutation relations:
\eq$$
\Aligned{
[\bfD_k,\bfD_l]
&=-2k\delta_{k+l,0},
\quad
[\bfD_k,d^+_l]=2,
\quad
k,l\in\bbDodd_s,
\\
[\bfd^\pm_k,\bfd^\mp_l]
&=-k\delta_{k+l,0},
\quad
k,l\in\bbDeven_{s,\ne0}.
}\label{bfd-commut}
$$
The usual elements $d^\pm_k$ for special values of $k$ are degenerate:
\eq$$
d^-_k=(-)^{k-1}d^+_k,
\qquad
d^+_k|1\rangle_a=0,
\qquad
{}_a\langle1|d^+_k=0
\quad\text{for}\quad
k\in\bbD_s.
\label{dpm-coincide}
$$
We omit explicit definitions of the currents in terms of these elements. They are cumbersome, though easily derived, and they will not be used directly. Operator products and commutation relations of currents are continued to the special values of~$r$ analytically. For later use we need the following commutation relations:
\Gather$$
[d^+_k,S_l]=0,
\quad
[d^+_k,S^+_l]=0,
\quad
[d^+_k,\eta_l]=0,
\label{d+-SS+eta-commut}
\\
[d^+_k,\tS_l]=-2\tS_{k+l},
\quad\text{if $k\in\bbDodd_s$}.
\label{d+-tS-commut}
$$
The first line of commutation relations holds due to the fact that $q^{k/2}-q^{-k/2}$ remains nonzero for~$k\in\bbDodd_s$ as $\tq\to\omega$, while $d^-_k-d^+_k$ in the numerator vanishes.

Another peculiarity of the special case~(\ref{rs-values}) is the identity
\eq$$
a_{mn}=a_{m+2,n+2s+1}
\quad\text{for $r={2s+1\over2s-1}$.}
\label{rs-amn-id}
$$
This leads to the identification of the corresponding spaces: $\cF_{mn}=\cF_{m+2,n+2s+1}$, $\bcF_{mn}=\bcF_{m+2,n+2s+1}$. Nevertheless, the screening operators $\Sigma$, $\tSigma$ are defined differently for different values of~$m,n$. We will distinguish them by the subscripts $mn$ at the vectors or by writing like $\cdots|_{\bcF_{mn}}$. The operators
\eq$$
\Sigma|_{\bcF_{m+2k,n+(2s+1)k}}=S_{m-n+1-(2s-1)k},
\qquad
\tSigma|_{\bcF_{m+2k,n+(2s+1)k}}=\tS_{n-m+1+(2s-1)k}
\label{tSigma-shifted}
$$
are different operators for different values of~$k$, though, in fact, they act on the same space~$\bcF_{mn}$.

Now let us turn to the reduction compatibility condition~(\ref{redcompat-J}). For the operators $V^{h\bh'}_{1n}$ it reads
\eq$$
C_s\prod^N_{i=1}h\left(z\over x_i\right)\>
{}_{1n}\langle h|t^{(2s-2)}(z)t(x_1)\cdots t(x_N)|h'\rangle_{1n}
={}_{1n}\langle h|t(z)t(x_1)\cdots t(x_N)|h'\rangle_{1n},
\label{redcondition}
$$
where the constant $C_s$ and the function $h(z)$ were defined in~(\ref{Ch-def}). Note that%
\footnote{This formula is derived by using the identity $(-1)^{s-1}\prod^{2s-2}_{j=1}\tg{\pi j\over2s-1}=\prod^{2s-2}_{j=1}{1-\omega^j\over1+\omega^j}=2s-1$.}
\eq$$
C_s=-K^{(2s-2)}/K^{(1)}.
\label{CKK-id}
$$
As we already mentioned, this condition seems to be very complicated for any treatment. In its explicit form the l.h.s.\ and r.h.s.\ expressions contain different numbers of terms, and it can be proved by means of the recursion relations in $N$ for exponential operators ($h=h'=1$) only. Here we propose a radically simpler and more general treatment by means of the currents~$\tau^{(\nu)}(z)$ defined in~(\ref{tau-def}).

Let us compare the currents $\tau^{(2s-2)}(z)$ and~$\tau^{(1)}(z)$. First, let $\tq=\omega\e^\epsilon$ with some small~$\epsilon$. Calculate the ratio $\lcolon\tau^{(1)}(z)(\tau^{(2s-2)}(z))^{-1}\rcolon$ and perform the limit $\epsilon\to0$. As a result we obtain
\eq$$
\tau^{(1)}(z)=\lcolon\zeta(z)\mu(z)\tau^{(2s-2)}(z)\rcolon.
\label{tau-tau-id}
$$
The factors $\zeta(z)$ and $\mu(z)$ are advisedly separated. The second one comes from the `regular' modes and is given by
\eq$$
\mu(z)=\lcolon\exp\sum_{k\ne0}{(-1)^k\omega^{k/2}-\omega^{-k/2}\over k}\bfD_kz^{-k}\rcolon,
\qquad
\bfD_k=\bfd^-_k+(-1)^k\bfd^+_k.
\label{mu-def}
$$
Here we extended the notation $\bfD_k$ to all values of~$k$. It explicitly commutes with $\tS_k$, since $\mu(z')\tS(z)=\lcolon\mu(z')\tS(z)\rcolon$. Its operator products read
\eq$$
\mu(z')t(z)=t(z)\mu(z')=h\left(z\over z'\right)\lcolon\mu(z')t(z)\rcolon
\qquad(z'\ne-z).
\label{mu-t-prod}
$$
It means that it only produces a product of the $h(z/x_i)$ functions being inserted into a matrix element of $t(X)$ between $\cA$\=/states.

The factor $\zeta(z)$ in (\ref{tau-tau-id}) is of different nature. It consists of just `dangerous' modes:
\eq$$
\zeta(z)=\lcolon\exp\sum_{k\in\bbDodd_s}{(2s-1)d^+_k\over k}z^{-k}\rcolon.
\label{zeta-def}
$$
It is convenient to explicitly split it into the positive and negative modes contributions:
\eq$$
\zeta(z)=\zeta^{(-)}(z)\zeta^{(+)}(z),
\qquad
\zeta^{(\pm)}(z)=\exp\sum_{k\in\pm\bbDodd_{s,>0}}{(2s-1)d^+_k\over k}z^{-k}.
\label{zetapm-def}
$$
The currents $\zeta^{(\pm)}(z)$ commute with $t(z')$, $s(z')$, but do not commute with the operators~$\tS_k$. With these currents eq.~(\ref{tau-tau-id}) looks like
$$
\tau^{(1)}(z)=\zeta^{(-)}(z)\lcolon\mu(z)\tau^{(2s-2)}(z)\rcolon\zeta^{(+)}(z).
$$
We want to get rid of $\zeta^{(\pm)}(z)$ in this equation. It is possible if we consider matrix elements between $\cA$\=/vectors that do not contain $c_{-k}$ with $k\in\bbDodd_s$.

Let us define the `$\cR$\=/subspaces' $\cF^\cR_{mn}$, $\bcF^\cR_{mn}$ as follows:
\eq$$
\Aligned{
\cF^\cR_{mn}
&=\Big\{{}_{mn}\langle h|\>\Big|\>h\in\check\cA,\ {}_{mn}\langle h|d^+_{-k}=0,\
{}_{mn}\langle h|\tSigma d^+_{-k}=0\text{ for }k\in\bbDodd_{s,>0}\Big\},
\\
\bcF^\cR_{mn}
&=\Big\{|h\rangle_{mn}\>\Big|\>h\in\check\cA,\ d^+_k|h\rangle_{mn}=0,\
d^+_k\tSigma|h\rangle_{mn}=0\text{ for }k\in\bbDodd_{s,>0}\Big\}.
}\label{R-spaces-def}
$$
In other words, these spaces consist of vectors such that neither the vector not the action of the $\tSigma$ operator of the vector contain any `dangerous' modes. Here such vectors will be called $\cR$\=/vectors. The operators $\zeta^{(\pm)}(z)$ act as unity on the vectors ${}_{mn}\langle h|$, ${}_{mn}\langle h|\tSigma$, $|h'\rangle_{mn}$, $\tSigma|h'\rangle_{mn}$, if ${}_{mn}\langle h|$ and $|h'\rangle_{mn}$ are $\cR$\=/vectors. Hence, we have
$$
\Aligned{
{}_{mn}\langle h|\tSigma\tau^{(1)}(z)t(X)|h'\rangle_{mn}
&={}_{mn}\langle h|\tSigma\lcolon\mu(z)\tau^{(2s-2)}(z)\rcolon t(X)|h'\rangle_{mn},
\\
{}_{mn}\langle h|\tau^{(1)}(z)\tSigma t(X)|h'\rangle_{mn}
&={}_{mn}\langle h|\lcolon\mu(z)\tau^{(2s-2)}(z)\rcolon\tSigma t(X)|h'\rangle_{mn}.
}
$$
Let $m=1$. Write down two of the identities~(\ref{tauS-commut}):
\eq$$
\Aligned{
\tau^{(1)}(z)\tSigma-z^2\tSigma\tau^{(1)}(z)|_{\bcF_{1n}}
&=K^{(1)}z^nt^{(1)}(z)|_{\bcF_{1n}},
\\
\lcolon\mu(z)\tau^{(2s-2)}(z)\rcolon\tSigma+z^2\tSigma\lcolon\mu(z)\tau^{(2s-2)}(z)\rcolon|_{\bcF_{1n}}
&=K^{(2s-2)}z^n\lcolon\mu(z)t^{(2s-2)}(z)\rcolon|_{\bcF_{1n}}.
}
\label{t1t2s-2-in-tau}
$$
By taking the sum of these two identities we obtain
\eq$$
{}_{1n}\langle h|\Delta(z)t(X)|h'\rangle_{1n}
=2B_1z^{-n}\>{}_{1n}\langle h|\tau^{(1)}(z)t(X)\tSigma|h'\rangle_{1n},
\quad
\text{if ${}_{1n}\langle h|\in\cF^\cR_{1n}$, $|h'\rangle_{1n}\in\bcF^\cR_{1n}$,}
\label{t1t2s-2-id}
$$
where
\eq$$
\Delta(z)=t^{(1)}(z)-C_s\lcolon\mu(z)t^{(2s-2)}(z)\rcolon.
\label{Delta-def}
$$
The r.~h.~s.\ of (\ref{t1t2s-2-id}) vanishes, if the right vector is killed by the screening operator and, hence,
\Multline$$
C_s\prod^s_{i=1}h\left(z\over x_i\right)\>{}_{1n}\langle h|t^{(2s-2)}(z)t(X)|h'\rangle_{1n}
={}_{1n}\langle h|t^{(1)}(z)t(X)|h'\rangle_{1n},
\\
\text{if ${}_{1n}\langle h|\in\cF^\cR_{1n}$, $|h'\rangle_{1n}\in\bcF^\cR_{1n}$ and $\tSigma|h'\rangle_{1n}=0$.}
\label{t1t2s-2-zeroid}
$$

It is easy to apply this condition to the exponential operators $V_{1n}(x)$. The vector $\tSigma|1\rangle_{1n}=\tS_n|1\rangle_{1n}$ vanishes if $n>0$. On the other hand, the vector ${}_{1n}\langle1|\tSigma={}_{1n}\langle1|\tS_{n-2}$ does not contain dangerous modes if $n-2<2s-1$. Hence, $1\le n\le 2s$ in consistency with the conjecture that they coincide with primary conformal fields~$\Phi_{1n}(x)$.

Similarly, we can derive the reduction compatibility conditions for the operators at the reflected point $a=-a_{1n}=a_{-1,-n}$. It is easy to check that
\eq$$
\Aligned{
\tau^{(1)}(z)\tSigma-z^2\tSigma\tau^{(1)}(z)|_{\bcF_{-1,-n}}
&=-K^{(1)}z^{2-n}t^{(1)}(z)|_{\bcF_{-1,-n}},
\\
\lcolon\mu(z)\tau^{(2s-2)}(z)\rcolon\tSigma+z^2\tSigma\lcolon\mu(z)\tau^{(2s-2)}(z)\rcolon|_{\bcF_{-1,-n}}
&=K^{(2s-2)}z^{2-n}\lcolon\mu(z)t^{(2s-2)}(z)\rcolon|_{\bcF_{-1,-n}},
}
\label{t1t2s-2-in-tau-reflected}
$$
Due to the minus sign in the r.h.s.\ of the first line we have to take the difference of these two equations:
\Multline$$
{}_{-1,-n}\langle h|\Delta(z)t(X)|h'\rangle_{-1,-n}
=-2B_1z^n\>{}_{-1,-n}\langle h|\tSigma\tau^{(1)}(z)t(X)|h'\rangle_{-1,-n},
\\
\text{if ${}_{-1,-n}\langle h|\in\cF^\cR_{-1,-n}$, $|h'\rangle_{-1,-n}\in\bcF^\cR_{-1,-n}$.}
\label{t1t2s-2-id-reflected}
$$
This results in the compatibility condition
\Multline$$
C_s\prod^s_{i=1}h\left(z\over x_i\right)\>{}_{-1,-n}\langle h|t^{(2s-2)}(z)t(X)|h'\rangle_{-1,-n}
={}_{-1,-n}\langle h|t^{(1)}(z)t(X)|h'\rangle_{-1,-n},
\\
\text{if ${}_{-1,-n}\langle h|\in\cF^\cR_{-1,-n}$, $|h'\rangle_{-1,-n}\in\bcF^\cR_{-1,-n}$ and ${}_{-1,-n}\langle h|\tSigma=0$.}
\label{t1t2s-2-zeroid-reflected}
$$
In fact, we could use the condition~(\ref{t1t2s-2-zeroid}), since $a_{-1,-n}=a_{1,2s+1-n}$ due to~(\ref{rs-amn-id}). But in the next section we will need this alternative form of the reduction compatibility condition.

\section{Local operators compatible with the reduction}
\label{locop-comp-sec}

The condition (\ref{t1t2s-2-zeroid}) that the operator $V^{h\bar h'}_{1n}(x)$ be consistent with the reduction is chirally asymmetric and too restrictive with respect to~$h'$: it demands the extra condition that $\tSigma|h'\rangle_{1n}=0$. The same can be said about the condition (\ref{t1t2s-2-zeroid-reflected}) for $V^{h\bh'}_{-1,-n}(x)$. Our aim is to modify the operators $V^{h\bh'}_{\pm1,\pm n}(x)$ in such a way that such extra conditions become unnecessary. To do it let us add the $a$\=/derivatives $V^{\prime h\bar h'}_a(x)={d\over da}V^{h\bar h'}_a(x)$ at particular values of~$a$. The resulting operators $V^{\cR\>h\bh'}_{\pm1,\pm n}(x)$ would be compatible with the reduction subject to the only conditions that the vectors corresponding to $h$ and $h'$ belong to the $\cR$\=/subspaces defined in~(\ref{R-spaces-def}).

In this section it will be proved that the operator $V^{\cR\>h\bh'}_{1n}(x)$ defined by the $J$\=/functions
\eq$$
J^{\cR\,h\bar h'}_{1n}(X)
={}_{1n}\langle h|t(X)|h'\rangle_{1n}
+{(-1)^{n-1\over2}\over\pi}\>{}_{1n}\langle h|[t'(X),\Sigma]\tSigma|h'\rangle_{3,n+2}
\quad\text{for $n\in2\Z+1$}
\label{JR-def}
$$
and the operators $V^{\cR\>h\bh'}_{-1,-n}(x)$ defined by the $J$\=/functions
\eq$$
J^{\cR\,h\bar h'}_{-1,-n}(X)
={}_{-1,-n}\langle h|t(X)|h'\rangle_{-1,-n}
+{(-1)^{n-1\over2}\over\pi}\>{}_{-3,-n-2}\langle h|\tSigma[t'(X),\Sigma]|h'\rangle_{-1,-n}
\quad\text{for $n\in2\Z+1$}
\label{JR-def-reflected}
$$
are compatible with the reduction if ${}_{1n}\langle h|\in\cF^\cR_{1n}$, $|h'\rangle_{1n}\in\bcF^\cR_{1n}$ or ${}_{-1,-n}\langle h|\in\cF^\cR_{-1,-n}$, $|h'\rangle_{-1,-n}\in\bcF^\cR_{-1,-n}$ respectively.

Eqs.~(\ref{JR-def}) and (\ref{JR-def-reflected}) define the operators $V^{\cR\>h\bh'}_{\pm1,\pm n}$ for odd values of $n$ only. Nevertheless, it is not too restrictive. Due to the identity~(\ref{rs-amn-id}) we have $a_{1n}=a_{-1,n-2s-1}$. The parities of $n$ and $n^*=2s+1-n$ are opposite. Hence, we may identify
\eq$$
V^{\cR\>h\bh'}_{1n}(x)=V^{\cR\>h\bh'}_{-1,-n^*}(x).
\label{VcR-oddeven-def}
$$
This makes it possible to define $V^{\cR\>h\bh'}_{\pm1,\pm n}(x)$ for $n$ even, as soon as we have a definition for $n$ odd.

To prove the main statement for the operator~$V^{\cR\>h\bh'}_{1n}$ it is necessary to show that
\eq$$
{}_{1n}\langle h|\Delta(z)t(X)|h'\rangle_{1n}
+{(-1)^{n-1\over2}\over\pi}{}_{1n}\langle h|[(\Delta(z)t(X))',\Sigma]\tSigma|h'\rangle_{3,n+2}=0.
\label{JR-Delta-zero}
$$
The first term here is given by (\ref{t1t2s-2-id}) and is generally nonzero. Let us transform the second term
\Multline*$$
{}_{1n}\langle h|[(\Delta(z)t(X))',S_{-n}]\tS_n|h'\rangle_{3,n+2}
={}_{1n}\langle h|[(\Delta(z)t(X))',\Sigma]\tSigma|h'\rangle_{3,n+2}
\\
={}_{1n}\langle h|[\Delta'(z),\Sigma]t(X)\tSigma|h'\rangle_{3,n+2}
+\sum^N_{i=1}{}_{1n}\langle h|\Delta(z)t(X_{<i})[t'(x_i),\Sigma]t(X_{>i})\tSigma|h'\rangle_{3,n+2}.
$$
Here $X_{<i}=(x_1,\ldots,x_{i-1})$, $X_{>i}=(x_{i+1},\ldots,x_N)$. Substituting (\ref{S't-mnodd-tau}) we obtain that in any matrix element between $\cR$\=/states
\Multline*$$
[\Delta'(z),\Sigma]_{\bcF_{1,n+2}}
=(-1)^{n+1\over2}\pi B_1z^{-n}\left(\epsilon_0\tau^{(1)}(z)+\tau^{(1)}(z)\epsilon_0
-\epsilon_0\lcolon\mu(z)\tau^{(2s-2)}(z)\rcolon+\lcolon\mu(z)\tau^{(2s-2)}(z)\rcolon\epsilon_0\right)
\\
=2\pi(-1)^{n+1\over2}z^{-n}\tau^{(1)}(z)\epsilon_0.
$$
Then we get
\Multline*$$
{}_{1n}\langle h|[(\Delta(z)t(X))',\Sigma]\tSigma|h'\rangle_{3,n+2}
=(-1)^{n+1\over2}2\pi B_1z^{-n}\>{}_{1n}\langle h|\tau^{(1)}(z)\epsilon_0t(X)\tSigma|h'\rangle_{3,n+2}
\\
+(-1)^{n+1\over2}\pi B_1\sum^N_{i=1}x_i^{-n}\>
{}_{1n}\langle h|\Delta(z)t(X_{<i})\left(\epsilon_0\tau^{(1)}(x_i)+\tau^{(1)}(x_i)\epsilon_0\right)
t(X_{>i})\tSigma|h'\rangle_{3,n+2}
\\
=(-1)^{n+1\over2}2\pi B_1z^{-n}\>{}_{1n}\langle h|\tau^{(1)}(z)\epsilon_0t(X)\tSigma|h'\rangle_{3,n+2}
\\
+(-1)^{n+1\over2}2\pi B_1^2z^{-n}\sum^N_{i=1}x_i^{-n}\>
{}_{1n}\langle h|\tau^{(1)}(z)\tSigma t(X_{<i})\left(\epsilon_0\tau^{(1)}(x_i)+\tau^{(1)}(x_i)\epsilon_0\right)
t(X_{>i})\tSigma|h'\rangle_{3,n+2}.
$$
Pushing the first $\tSigma$ in the last line to the right and taking into account that $\tSigma^2=0$, we obtain
\Multline*$$
{}_{1n}\langle h|[(\Delta(z)t(X))',\Sigma]\tSigma|h'\rangle_{3,n+2}
=(-1)^{n+1\over2}2\pi B_1z^{-n}\>{}_{1n}\langle h|\tau^{(1)}(z)\epsilon_0t(X)\tSigma|h'\rangle_{3,n+2}
\\
-(-1)^{n+1\over2}2\pi B_1z^{-n}\sum^N_{i=1}
{}_{1n}\langle h|\tau^{(1)}(z)t(X_{<i})\left(\epsilon_0t(x_i)-t(x_i)\epsilon_0\right)
t(X_{>i})\tSigma|h'\rangle_{3,n+2}
\\
=(-1)^{n+1\over2}2\pi B_1z^{-n}
\left(
{}_{1n}\langle h|\tau^{(1)}(z)\epsilon_0t(X)\tSigma|h'\rangle_{3,n+2}
+{}_{1n}\langle h|\tau^{(1)}(z)[t(X),\epsilon_0]\tSigma|h'\rangle_{3,n+2}
\right).
$$
After summing up two last terms, the operator $\epsilon_0$ only remains to the right of $t(X)$ and only shifts the zero mode:
\eq$$
{}_{1n}\langle h|[(\Delta(z)t(X))',\Sigma]\tSigma|h'\rangle_{3,n+2}
=(-1)^{n+1\over2}2\pi B_1z^{-n}\>{}_{1n}\langle h|\tau^{(1)}(z)t(X)\tSigma|h'\rangle_{1n}.
\label{V'hh'-delta}
$$
We see that the r.h.s.\ of this equation coincides (up to a constant factor) with the r.h.s.\ of~(\ref{t1t2s-2-id}). Thus we obtain (\ref{JR-Delta-zero}), which proves the compatibility with the reduction of~$V^{\cR\>h\bh'}_{1n}(x)$.

Analogously, for $a=a_{-1,-n}$ the identity
$$
{}_{-1,-n}\langle h|\Delta(z)t(X)|h'\rangle_{-1,-n}
+{(-1)^{n-1\over2}\over\pi}{}_{-3,-n-2}\langle h|\tSigma[(\Delta(z)t(X))',\Sigma]|h'\rangle_{-1,-n}=0
$$
is derived by using (\ref{t1t2s-2-id-reflected}), which proves the compatibility with the reduction of the operator $V^{\cR\,hh'}_{-1,-n}(x)$.

\section{Creating \texorpdfstring{$\cR$}{R}-vectors by means of the \texorpdfstring{$S^+_k$}{S+(k)} and \texorpdfstring{$\tS_k$}{tilde S(k)} modes}
\label{S+-sec}

Among the vectors (\ref{scr-vectors}) the $\cR$\=/vectors could be obtained by means of the elements $\ft^+_{-k_1,\ldots,-k_M}$ and~$\eta_{-k}$ due to the commutation relations~(\ref{d+-SS+eta-commut}) and~(\ref{StS+-commut}). The elements $\eta_{-k}$ being combinations of odd\-/level generating elements of $\check\cA$ (which generate, as we have said, integrals of motion) provide a trivial contribution to form factors and we will ignore them, keeping in mind that if any vector ${}_{1n}\langle h|$ is an $\cR$\=/vector, the vector ${}_{1n}\langle h\eta_{-k}|$ is an $\cR$\=/vector as well.

Now turn our attention to the vectors ${}_{1n}\langle\ft^+_{-k_1,\ldots,-k_M}|$. They can be $\cR$\=/vectors, but subject to very restrictive condition on the value of~$M$. Indeed, this vector evidently satisfies the first $\cR$\=/vector condition: ${}_{1n}\langle\ft^+_{-k_1,\ldots,-k_M}|d^-_{-k}=0$, if $k\in\bbDodd_{s,>0}$. But it is not so obvious concerning the second $\cR$\=/vector condition. By using (\ref{d+-tS-commut}) we obtain
$$
{}_{1n}\langle\ft^+_{-k_1,\ldots,-k_M}|{1\over2}\tSigma d^+_{-k}
={}_{1,n+2M}\langle1|S^+_{k_1}\cdots S^+_{k_M}\tS_{n-2-k}
={}_{1,n+2M}\langle1|\tS_{n+2M-2-k}S^+_{k_1-2}\cdots S^+_{k_M-2}.
$$
The r.h.s.\ vanishes, if $n+2M-2-k<0$. The minimal value of $k$ that enters the $\cR$\=/vector condition is~$2s-1$. Hence, if
$$
n+2M\le2s,
$$
the vector ${}_{1n}\langle\ft^+_{-k_1,\ldots,-k_M}|$ is an $\cR$\=/vector. If, for example, $n=2s-1$ the only solution is $M=0$, so that we remain with the vacuum vector (modulo the action of integrals of motion) only. The space of $\cR$\=/vectors generated by the elements $\ft^+_{-k_1,\ldots,-k_M}$ is definitely too small to describe the whole space of states of the perturbed minimal model.

To solve the problem, let us try to insert the operators $\tS_k$ into the expression for ${}_{1n}\langle\ft^+_{-k_1,\ldots,-k_M}|$, but accurately, so that the `dangerous' modes would not appear. Consider the vectors
\eq$$
{}_{1n}\langle\ft^{+(p,n)}_{-k_1,\ldots,-k_M}|
={}_{1-2p,n+2M}\langle1|\prod^{\substack{\curvearrowright\\M}}_{i=1}S^+_{k_i}
\prod^{\substack{\curvearrowleft\\p}}_{j=1}\tS_{n-2-j(2s-3)}.
\label{t+mod-bra-def}
$$
It can be checked (see Appendix~\ref{Rvec-appendix}) that these vectors are $\cR$\=/vectors subject to the condition
\eq$$
\Gathered{
n+2M\le n_1
\quad\text{for $p=0$,}
\\
n_p\le n+2M\le n_{p+1}
\quad\text{for $p>0$,}
}\label{t+mod-R-bra-condition}
$$
where
\eq$$
n_p=p(2s-3)+3.
\label{np-def}
$$
The lower condition $n+2M\ge n_p$ is in fact not an $\cR$\=/vector condition, but a kind of `nontriviality' condition. For $n+2M<n_p-1$ the vector vanishes. In the case $n+2M=n_p-1$ it reduces to the $(p-1)$\=/case.

Note that the operators $\tS_{n-2-j(2s-3)}$ commute with the current~$t(z)$. Indeed,
$$
\tS_{n-2-j(2s-3)}=\tSigma|_{\bcF_{3-4j,n-j(2s+1)}},
$$
but $a_{3-4j,n-j(2s+1)}=a_{1-2(j-1),n}$ is the correct value of $a$ to the right of this operator in eq.~(\ref{t+mod-bra-def}). Since the first identity of~(\ref{tS+tS-commut}) can be rewritten as $\tS_k\tSigma=\tSigma\tS_{k-2}$, each of the operators $\tS_{n-2-j(2s-3)}$ being pulled to the right of all the $\tS_k$ modes in (\ref{t+mod-bra-def}) remains to be the $\tSigma$ operator. Explicitly, it becomes $\tS_{n-j(2s-1)}$. The subscript of this operator must be nonpositive for any $j=1,\ldots,p$ so that it would not make the vector (\ref{t+mod-bra-def}) the null vector. Hence, this vector is sensible for $n\le2s-1$.

The analogous ket-vectors are
\eq$$
|\ft^{+(p,2-n)}_{-k_1,\ldots,-k_M}\rangle_{1n}=\prod^{\substack{\curvearrowright\\p}}_{j=1}\tS_{n+j(2s-3)}
\prod^{\substack{\curvearrowleft\\M}}_{i=1}S^+_{-k_i}|1\rangle_{1+2p,n-2M}.
\label{t+mod-ket-def}
$$
The corresponding $\cR$\=/vector condition is
\eq$$
\Gathered{
2M-n+2\le n_1
\quad\text{for $p=0$,}
\\
n_p\le 2M-n+2\le n_{p+1}
\quad\text{for $p>0$.}
}\label{t+mod-R-ket-condition}
$$
Both vectors (\ref{t+mod-bra-def}) and (\ref{t+mod-ket-def}) are not null vectors if
\eq$$
|n|\le2s-1.
\label{n-restriction}
$$
Hence, these conditions (\ref{t+mod-R-bra-condition}) and (\ref{t+mod-R-ket-condition}) are applicable for building the operators $V^{\cR\,hh'}_{1n}(x)$ for~$n$ odd, $n=1,3,\ldots,2s-1$. For $n$ even ($n=2,4,\ldots,2s$) we have to substitute $n$ by $n^*$ and exchange the conditions for bra- and ket-vectors.

\section{Conserved currents compatible with the reduction}
\label{conscur-sec}

Here we discuss an important example of local operators: the conserved currents $T_{2k}$, $\Theta_{2k-2}$, $T_{-2k}$, $\bar\Theta_{2-2k}$ ($k=1,2,\ldots$) compatible with the reduction. The conserved currents are densities of the commutative integrals of motion:
\eq$$
\Aligned{
I_{2k-1}
&=\int{dz\over2\pi}\,T_{2k}(x)+\int{d\bz\over2\pi}\,\Theta_{2k-2}(x),
\\
I_{1-2k}
&=\int{dz\over2\pi}\,\Theta_{2-2k}(x)+\int{d\bz\over2\pi}\,T_{-2k}(x).
}\label{IM-def}
$$
Here the integration contour is any space-like line. When it is the line $x^0=\const$ both differentials are equal: $dz=d\bz=dx^1$. The subscripts at conserved charges and currents denote the (Lorentz) spin of the operators.

In terms of currents the conservation laws are expressed as continuity equations:
\subeq{\label{continuity}
\Align$$
\bd T_{2k}
&=\d\Theta_{2k-2},
\label{continuity-posspin}
\\
\d T_{-2k}
&=\bd\Theta_{2-2k}.
\label{continuity-negspin}
$$}
The currents $T\equiv T_2=-2\pi T_{zz}$, $\bT\equiv T_{-2}=-2\pi T_{\bz\bz}$ and $\Theta\equiv\Theta_0=2\pi T_{z\bz}$ are proportional to the components of the energy\-/momentum tensor~$T_{\mu\nu}$.

The conserved currents are defined by these equations not uniquely, since we may admix to every current a multiple commutator of integrals of motion with another current. This operation may change the set of integrals of motion, but not the spectrum of their eigenvalues. In~\cite{Lashkevich:2013yja} we proposed the currents $T_{2k+2}$, $\Theta_{2k}$ in the form
$$
T^{\text{\!\!\cite{Lashkevich:2013yja}}}_{2k}(x)={\i\pi m^{2k}\over8B_2}V^{\ft_{-2k}}_{11}(x),
\qquad
\Theta^{\!\!\text{\cite{Lashkevich:2013yja}}}_{2k-2}(x)=-{2\pi\mu m^{2k-2}\over r}G_{13}V^{\ft_{2-2k}}_{13}(x),
$$
and $T_{-2k}$, $\Theta_{2-2k}$ in a similar form. The normalization is chosen in such a way that for $k=1$ it provides the correct normalization of the energy\-/momentum tensor, while for general $k$ it only provides the correct dimensionality. These operators were shown to satisfy the continuity equations. Nevertheless, a more detailed study shows that such conserved currents are not compatible with the reduction. Indeed, the vectors ${}_{11}\langle\ft_{-2k}|$, ${}_{13}\langle\ft_{2-2k}|$ are not $\cR$\=/vectors. Here we use the above\-/mentioned non\-/uniqueness to define another set of conserved currents. To do it, take into account that the vectors ${}_{11}\langle\ft^+_{-2k}|$ and $|\ft^+_{-2k}\rangle_{11}$, being vectors of the form~(\ref{t+mod-bra-def}) and~(\ref{t+mod-ket-def}), are $\cR$\=/vectors. Then by using eqs.~(\ref{JR-def}), (\ref{JR-def-reflected}) we construct conserved currents compatible with the reduction. For the positive spin currents we have
\eq$$
\Aligned{
T_{2k}(x)
&=-{\i\pi m^{2k}\over8B_2}V^{\cR\,\ft^+_{-2k}}_{11}(x)
=-{\i\pi m^{2k}\over8B_2}V^{\ft^+_{-2k}}_{11}(x),
\\
\Theta_{2k-2}(x)
&=-{2\pi\mu m^{2k-2}\over r}G_{13}V^{\cR\,\ft^+_{2-2k}}_{-1,-3}(x)
=-{2\pi\mu m^{2k-2}\over r}G_{13}V^{\ft^+_{2-2k}}_{-1,-3}(x).
}\label{TTheta-posspin-def}
$$
The negative spin currents are defined similarly, but the explicit expression is more complicated, since it really contains an $a$\=/derivative:
\eq$$
\Aligned{
T_{-2k}(x)
&=-{\i\pi m^{2k}\over8B_2}V^{\cR\,\bar\ft^+_{-2k}}_{11}(x)
=-{\i\pi m^{2k}\over8B_2}\left(V^{\bar\ft^+_{-2k}}_{11}(x)
-{\tB_1\over2\pi}V^{\prime\,\overline{\eta^{\vphantom{1}}_{1-2k}c_{-1}}}_{11}(x)\right),
\\
\Theta_{2-2k}(x)
&=-{2\pi\mu m^{2k-2}\over r}G_{13}V^{\cR\,\bar\ft^+_{2-2k}}_{13}(x)
=-{2\pi\mu m^{2k-2}\over r}G_{13}V^{\bar\ft^+_{2-2k}}_{13}(x).
}\label{TTheta-negspin-def}
$$
In the case $k=1$ it coincides with the known expression obtained in~\cite{Feigin:2008hs} (see (A.17) there). The operators defined here satisfy the continuity equations~(\ref{continuity}). The proof of this fact can be found in Appendix~\ref{continuity-negspin-appendix}.

Note that though the reduction procedure only makes sense for special values of~$r$, the conserved currents (\ref{TTheta-posspin-def}) and (\ref{TTheta-negspin-def}), as well as the general operators $V^{\cR\>h\bh'}_{\pm1,n}$ ($n$ odd), are well-defined for arbitrary $r>0$. We expect that there is another, more deep, reason for these expressions, like the `charge at infinity' picture in conformal field theory~\cite{Dotsenko:1984nm}.

There is another question regarding conserved currents at $r={2s+1\over2s-1}$. Integrals of motion $I_\sigma$ ($\sigma$ being odd integer $\pm(2k-1)$) of these model are known~\cite{Smirnov:1990vm,Eguchi:1989dq} to skip the values of the spin $\sigma$ divisible by $2s-1$. But the construction described here provides currents with every even value of spin. How is it possible? We may expect that the operators $I_\sigma$ for $\sigma\in\bbDodd_s$ can be expressed in terms of other integrals of motion. That is, the currents $T_{\sigma+1}$, $\Theta_{\sigma-1}$ (if $\sigma>0$) and $T_{\sigma-1}$, $\Theta_{\sigma+1}$ (if $\sigma<0$) for these values of spin should be expressed in terms of commutators of integrals of motion with lower currents (i.e.\ currents with smaller values of $|\sigma|$). Let us make sure that this is the case. It will be easier to derive this property in terms of the $\Theta$\=/components. The corresponding property for the $T$\=/components then follows from the continuity equations~(\ref{continuity}). We produce the derivation for the current $\Theta_{\sigma-1}\sim V^{\ft^+_{1-\sigma}}_{-1,-3}$ with $\sigma>0$. The derivation for $\Theta_{\sigma+1}\sim V^{\bar\ft^+_{1+\sigma}}_{13}$ with $\sigma<0$ is quite analogous.

Introduce the elements $\tft'_{-k}$ according to the rule (compare it with (\ref{ftk-def})):
\eq$$
\sum_{k\ge0}\tft'_{-k}z^k
=\exp\sum_{k\in\Z_{>0}\setminus\bbDodd_s}\tB_kc_{-k}z^k
=\sum_{k\ge0}\tft_{-k}z^k\times\exp\sum_{k\in\bbDodd_{s,>0}}(-\tB_k)c_{-k}z^k.
\label{tildeft'-def}
$$
Any element $\tft'_{-k}$ is a linear combination of the elements of the form $\tft'_{-k+\sum l_i}\prod c_{-l_i}$, $l_i\in\bbDodd_s$, and does not contain dangerous elements $c_{-l}$, $l\in\bbDodd_s$, in its expansion into products of the generators~$c_{-j}$. We shall be interested in the particular elements $\tft'_{1-\sigma}$ with $\sigma\in\bbDodd_{s,>0}$. These elements possess the property that the corresponding vectors ${}_{-1,-3}\langle t'_{1-\sigma}|$ are linear combinations of the vectors ${}_{-1+2k,-3+k(2s-1)}\langle\,\prod c_{-l_i}|\tSigma$, where $k(2s-1)=\sigma-\sum l_i$. Hence,
$$
{}_{-1,-3}\langle\tft'_{1-\sigma}|t(X)|1\rangle_{-1,-3}=0
\quad\text{for $\sigma\in\bbDodd_s$.}
$$
It means that $V^{\tft'_{1-\sigma}}_{-1,-3}(x)=0$.

Note that
\eq$$
B_{2k}=(-1)^{k-1}\tB_{2k},
\quad\text{if $r={2s+1\over2s-1}$.}
\label{B2k-id}
$$
It results in the fact that the difference $\ft^+_{1-\sigma}-\tft'_{1-\sigma}$ ($\sigma\in\bbDodd_{s,>0}$) vanishes on the factor of $\check\cA$ over the ideal generated by the elements~$c_{1-2k}$. In the whole algebra $\check\cA$ it is a linear combination of terms, which contain at least one element $c_{-j}$ with $j\in(2\Z+1)\setminus\bbDodd_s$ each. By using (\ref{B2k-id}) it can be shown that, in fact, the terms are necessarily grouped into elements~$\ft^+_{-j}\prod_ic_{-j_i}$ with $j$ even and all $j_i$ odd. Hence, the operator $\Theta_{\sigma-1}$ is a linear combination of commutators of integrals of motion with lower conserved currents, as we expected.

Following the reduction construction we can now propose a representation for local products of a right $T_{2k}$ and a left $T_{-2l}$ chiral currents:
\Multline$$
(T_{2k}T_{-2l})(x)
=-{\pi^2m^{2k+2l}\over64B_2^2}V^{\cR\,\ft^+_{-2k}\bar\ft^+_{-2l}}_{11}(x)
\\
=-{\pi^2m^{2k+2l}\over64B_2^2}\left(
  V^{\ft^+_{-2k}\bar\ft^+_{-2l}}_{11}(x)-{\tB_1\over2\pi}\left(
    V^{\prime\,\ft^+_{-2k}\overline{\eta^{\vphantom{1}}_{1-2l}c_{-1}}}_{11}(x)
    -V^{\prime\,\eta_{1-2k}\overline{\ft^+_{2-2l}c_{-1}}}_{13}(x)
  \right)
\right).
\label{TbarT-gen-def}
$$
The resulting operator is a level $(2k,2l)$ descendant operator in the family of the unit operator. Recall that this equation uniquely defines the whole set of exact form factors of the operator as follows. Eqs.~(\ref{screl-series}) and (\ref{Bk-def}) make it possible to express the algebraic elements $\ft^+_{2k}$ etc.\ in terms of the generators $c_{-k}$ of the algebra~$\check\cA$. Then eqs.~(\ref{c-bc-commut})--(\ref{Jhh'-explicit}) together with (\ref{FJ-rel}) define the first\-/breather form factors. Form factors containing other breathers are obtained by using the fusion procedure~(\ref{F-fusion}).

The simplest case $T\bT(z)$ ($k=l=1$) has been discussed in the literature, and so we write it down more explicitly:
\Multline$$
T\bT(x)
=-{\pi^2m^4\over64}\biggl(
  V^{c_{-2}\bc_{-2}}_{11}(x)-{2\i\over\pi}V^{\prime c_{-2}\bc_{-1}^2}_{11}(x)
\\
  -{B_1^2\over2B_2}\left(
    V^{c_{-2}\bc_{-1}^2+c_{-1}^2\bc_{-2}}_{11}(x)-{2\i\over\pi}V^{\prime c_{-1}^2\bc_{-1}^2}_{11}(x)\right)
  -{2\i\over\pi B_2}V^{\prime c_{-1}\bc_{-1}}_{13}(x)\biggr).
\label{TbarT-def}
$$
The operator $T\bT(x)$ was introduced by A.~Zamolodchikov~\cite{Zamolodchikov:2004ce} as a limit of the difference $T(x')\bT(x)-\Theta(x')\Theta(x)$ as $x'\to x$. This limit is well\-/defined in terms of the operator product expansions modulo partial derivatives in~$x$. A.~Zamolodchikov proved that its vacuum expectation value in any reasonable theory is given by $\langle T\bT\rangle=-\langle\Theta\rangle^2$. By using (\ref{Pc2barc2}) we obtain
$$
\langle T\bT\rangle=-{\pi^2m^4\over64\sin^2\pi r}.
$$
By comparing with the vacuum expectation value of $\langle\Theta\rangle$ from \cite{Fateev:1997yg} we make sure that Zamolodchikov's equation is satisfied.

For perturbed $M(2,2s+1)$ models the form factors  of the operator $T\bar{T}$ up to 9\=/particle case  were proposed in the series of papers by G.~Delfino and G.~Niccoli\cite{Delfino:2004vc,Delfino:2005wi,Delfino:2006te}. Their expressions for form factors from zero up to three particles in the case of the perturbed $M(2,5)$ (Lee--Yang) model were used by V.~Belavin and O.~Miroshnichenko~\cite{Belavin:2005xg} to find long-range expansions of two-point correlation functions $\langle T\bT(x)\Theta(0)\rangle$ and $\langle T\bT(x)T\bT(0)\rangle$. These expansions were found to be smoothly matching the short-range expansions based on conformal perturbation theory, pointing to correctness of the expressions for form factors of~\cite{Delfino:2004vc}. Delfino and Niccoli were searching the $T\bT$ operator as a linear combination of five operators: $1$, $\Theta$, $\d\bd\Theta$, $\d^2\bd^2\Theta$ and the extra operator associated with so\-/called kernel solution. Starting from Zamolodchikov's definition and implying an additional asymptotic condition, the cluster factorization property, they solved the form factor equations and found that the solution is unique, except for the coefficient at the resonance term~$\d\bd\Theta$, which is arbitrary.

Our expressions for form factors of the operator $T\bT$ differ from those of~\cite{Delfino:2005wi,Delfino:2006te}, except for the Lee--Yang case. We compared form factors up to four particles and found them differ by terms proportional to form factors of the operator $\d\bd\Phi_{15}$. We believe that the operators $\Phi_{1n}$, $n=5,7,\ldots$, which are absent in the Lee--Yang case, should be added to the Ansatz used in \cite{Delfino:2006te}. It especially concerns the operator $\d\bd\Phi_{15}$, since it appears in the operator product expansion of $\Theta(x')\Theta(x)$. However, we need a more precise definition than that of~\cite{Zamolodchikov:2004ce} to distinguish between our proposal and that of Delfino and Niccoli.

\section{Conclusion}
\label{conclusion-sec}

In this paper we considered the reduction of the sine\-/Gordon model to perturbed minimal models of the `ribbon' series $M(2,2s+1)$. We challenged the compatibility condition (\ref{redcompat}) and found that it can be reduced to finding elements of the spaces $\cF^\cR_{mn}$, $\bcF^\cR_{mn}$ defined in~(\ref{R-spaces-def}). We have shown that there is a way to construct a local operator compatible with the reduction for every pair of vectors in these spaces. A set of vectors in these spaces has been proposed, though we do not know, if it provides a complete basis. Form factors of conserved currents $T_{\pm2k}$, $\Theta_{\pm(2k-2)}$ and the products $T_{2k}T_{-2l}$ have been found. Our result for $T\bar{T}=T_2T_{-2}$ was compared with those of~\cite{Zamolodchikov:2004ce,Delfino:2004vc,Delfino:2006te}. Our formula (\ref{TbarT-gen-def}) has the advantage that it provides an answer for any number of particles and for arbitrary spins of currents. Recently, we have learned from F.~Smirnov that form factors of the operators $T_{2k}T_{-2k}$, including those that contain kinks, can also be obtained from the results of~\cite{Jimbo:2011bc}. It would be interesting to compare the results.

An important problem, which remains unsolved above, is the problem of completeness of the proposed space of physical operators compatible with the reduction. We hope that, with the explicit algebraic prescription we have got, this problem can be treated algebraically. Another interesting direction of study, which can be considered in connection with the proposed screening algebra construction, is a generalization of the analysis of form factors of descendant operators in the $Z_N$ Ising models\cite{Fateev:2006js,Fateev:2009kp} and in the $\Phi_{12}$ perturbations of the minimal models, including the Ising model in magnetic field\cite{Alekseev:2011my,Alekseev:2012jd}.

\section*{Acknowledgments}

We are grateful to M.~Bershtein, G.~Delfino, F.~Smirnov, A.~Zamolodchikov for discussions. Y.P.\ is thankful to SISSA and to G.~Mussardo and G.~Delfino for their hospitality during his visit in autumn 2013. The study of the conserved currents was supported by the Russian Science Foundation under the grant \#~14-12-01383. Other parts of the study were supported by the Russian Foundation of Basic Research under the grant \#~13-01-90614.

\Appendix

\section{Some reference data}
\label{reference-appendix}

The constant $\rho$ and the function $R(\theta)$, which enter the form factors according to~(\ref{FJ-rel}), are~\cite{Lukyanov:1997bp}
\eq$$
\Gathered{
\rho=\i(R(-\i\pi)\sin\pi(r-1))^{-1/2}
=\i\left(2\sin{\pi(r-1)\over2}\right)^{-1/2}\exp\int^{\pi(1-r)}_0{dt\over2\pi}\,{t\over\sin t},
\\
R(\theta)=\exp\left(
-4\int^\infty_0{dt\over t}\,
{\sh{\pi t\over2}\sh{\pi(1-r)t\over2}\sh{\pi rt\over2}
\over\sh^2\pi t}\ch(\pi-\i\theta)t
\right).
}\label{R-def}
$$
Since we are interested in the case $r>1$, the constant $\rho$ is purely imaginary. We choose the value of the root by the condition $\Im\rho>0$.

The constant $\rho^{(\nu)}$ is given by
\eq$$
\rho^{(\nu)}=\rho^\nu R_*^{\nu-1}{\prod^{\nu-1}_{j=2}R^{\nu-j}(\i\pi(1-r)j)\over\prod^{\nu-1}_{j=1}\Gamma^{(j)}},
\qquad
R_*=-\i\Res_{\theta=\i\pi(1-r)}R(\theta)=-{\tg\pi r\over R(-\i\pi r)}.
\label{rho(nu)-def}
$$
The function $R^{(\nu)}(\theta)$ is given by
\eq$$
R^{(\nu)}(\theta)=\prod^\nu_{j=1}R\left(\theta+{\i\pi(1-r)\over2}(\nu+1-2j)\right).
\label{R(nu)-def}
$$
It can be shown that for $r={2s+1\over2s-1}$ we have
\eq$$
{\rho^{(2s-2)}\over\rho}=C_s,
\qquad
{R^{(2s-2)}(\theta)\over R(\theta)}=h(\e^\theta)
\label{rhoR-ratios}
$$
with $C_s$ and $h(z)$ defined in~(\ref{Ch-def}).

The elements $\ft_{-k_1,\ldots,-k_M}$, $\ft^+_{-k_1,\ldots,-k_M}$, $\eta_{-k}$, $\epsilon_{-k}$ are given by the series
\eq$$
\Aligned{
\sum^\infty_{k_1=0}\sum^\infty_{k_2=-2}\cdots\sum^\infty_{k_M=2-2M}\ft_{-k_1,\ldots,-k_M}\prod^M_{i=1}z_i^{k_i}
&=\prod^M_{i<j}\left(1-{z_j^2\over z_i^2}\right)\exp\sum^\infty_{k=1}B_kc_{-k}\sum^M_{i=1}z_i^k,
\\
\sum^\infty_{k_1=0}\sum^\infty_{k_2=-2}\cdots\sum^\infty_{k_M=2-2M}\ft^+_{-k_1,\ldots,-k_M}\prod^M_{i=1}z_i^{k_i}
&=\prod^M_{i<j}\left(1-{z_j^2\over z_i^2}\right)\exp\sum^\infty_{k=1}(-B_k)c_{-k}\sum^M_{i=1}z_i^k,
\\
\span
\sum^\infty_{k=0}\eta_{-k}z^k
=\exp\sum^\infty_{\substack{k=1\\k\in2\Z+1}}(-2B_k)c_{-k}z^k,
\quad
\sum^\infty_{k=0}\epsilon_{-k}z^k
=\exp\sum^\infty_{\substack{k=1\\k\in2\Z+1}}2c_{-k}z^k.
}\label{screl-series}
$$
Among the elements $\ft_{-k_1,\ldots,-k_M}$, $\ft^+_{-k_1,\ldots,-k_M}$ only the elements with $k_{i+1}\ge k_i-1$, $\sum^M_{i=1}k_i>0$ are independent. The elements $\tilde\ft_{-k_1,\ldots,-k_M}$, $\tilde\ft^+_{-k_1,\ldots,-k_M}$ and $\teta_{-k}$ are defined analogously with the substitution of $B_k$ to~$\tB_k$.

\section{Explicit rules for \texorpdfstring{$J$}{J}-functions}
\label{Jfunc-appendix}

Let $\cA$ and $\bcA$ be two copies of the algebra $\cA$, but the generators of $\bcA$ will be denoted as~$\bc_{-k}$. Let us construct the algebra $\cA^2$ generated by elements $c_{-k}$, $\bc_{-k}$ so that
\eq$$
[c_{-k},\bc_{-l}]
=-(1+(-1)^k)k(A^+_k)^{-1}\delta_{kl}.
\label{c-bc-commut}
$$
To any element $g\in\cA^2$ we associate a polynomial $P^g(X|Y)$ of two sets of variables $X=(x_1,\ldots,x_N)$ and $Y=(y_1,\ldots,y_M)$ according to the rules
\eq$$
\Gathered{
P^1(X|Y)=1;
\\
P^{c_{-k}}(X|Y)=p_k(X)+(-)^{k-1}p_k(Y);
\qquad
P^{\bc_{-k}}(X|Y)=p_{-k}(Y)+(-)^{k-1}p_{-k}(X);
\\
P^{k_1g_1+k_2g_2}(X|Y)=k_1P^{g_1}(X|Y)+k_2P^{g_2}(X|Y),
\quad
k_1,k_2\in\C,\ g_1,g_2\in\cA^2;
\\
P^{hh'}(X|Y)=P^h(X|Y)P^{h'}(X|Y),
\quad
h,h'\in\cA;
\\
P^{\bh\bh'}(X|Y)=P^\bh(X|Y)P^{\bh'}(X|Y),
\quad
\bh,\bh'\in\bcA;
\\
P^{\bh'h}(X|Y)=P^h(X|Y)P^{\bh'}(X|Y),
\quad
h\in\cA,\ \bh'\in\bcA.
}\label{P-rules}
$$
Here $p_k(X)=\sum_ix_i^k$ are the Newton symmetric polynomials. Let us stress the order $\bh'h$ in the last line. To render it to the `normal' order $h\bh'$ we have to apply the rule~\ref{c-bc-commut}.

Then $J$\=/functions are given by
\eq$$
J^{h\bh'}_{\nu,N}(X)=\sum_{X=X_-\sqcup X_+}\e^{\i\pi a(\#X_--\#X_+)}P^{h\bh'}(X_-|X_+)
\prod_{x\in X_-\atop y\in X_+}f\left(x\over y\right).
\label{Jhh'-explicit}
$$
The sum is taken over all partitions of the (multi)set $X$ into two nonintersecting subsets $X_-$ and~$X_+$. The sign $\#X$ denotes the cardinal number of~$X$.

Notice that noncommutativity of the elements $c_{-2k}$ and $\bc_{-2k}$ due to~(\ref{c-bc-commut}) leads to a subtlety in calculation of $J$\=/functions: the functions $P^{h\bh'}(X|Y)$ do not factorize into the `right-mover' and `left-mover' parts. For example,
\eq$$
P^{c_{-2}\bc_{-2}}(X|Y)=P^{c_{-2}}(X|Y)P^{\bc_{-2}}(X|Y)-{4\over(q-q^{-1})^2}.
\label{Pc2barc2}
$$
This `nonfactorization' property, though looking absolutely artificial in these explicit formulas, is strongly dictated by the free field representation. It leads to a number of properties essentially used in the present article. In particular, it provides the correct vacuum expectation value of the operator $T\bT$.

\section{Proof of the commutation relations for screening operators}
\label{screening-rel-appendix}

The commutation relations (\ref{screening-commut}) follow from the following operator products:
\subeq{
\label{screening-OP}
\Align$$
S(z')S(z)
&=-{z^2\over z^{\prime2}}S(z)S(z')=\left(1-{z^2\over z^{\prime2}}\right)\lcolon S(z')S(z)\rcolon,
\label{SS-OP}
\\
S^+(z')S^+(z)
&=-{z^2\over z^{\prime2}}S^+(z)S^+(z')=\left(1-{z^2\over z^{\prime2}}\right)\lcolon S^+(z')S^+(z)\rcolon,
\label{S+S+-OP}
\\
S^+(z')S(z)
&=-{z^{\prime2}\over z^2}S(z)S^+(z')={1\over1-z^2/z^{\prime2}}\lcolon S^+(z')S(z)\rcolon,
\label{S+S-OP}
\\
S(z')\tS^+(z)
&={z^2\over z^{\prime2}}\tS^+(z)S(z')=\left(1+{z^2\over z^{\prime2}}\right)\lcolon S(z')\tS^+(z)\rcolon,
\label{StS+-OP}
\\
S(z')\tS(z)
&={z^{\prime2}\over z^2}\tS(z)S(z')={1\over1+z^2/z^{\prime2}}\lcolon S(z')\tS(z)\rcolon,
\label{StS-OP}
\\
S^+(z')\tS^+(z)
&={z^{\prime2}\over z^2}\tS^+(z)S^+(z')={1\over1+z^2/z^{\prime2}}\lcolon S^+(z')\tS^+(z)\rcolon.
\label{S+tS+-OP}
$$}
Besides, four more equations are immediately obtained by simultaneous adding/removing tildes in (\ref{SS-OP})--(\ref{StS+-OP}). The right-hand sides in eqs.~(\ref{screening-commut}) come from the residues of the poles, which are proportional to the following operators:
\subeq{
\label{OP-redidues}
\Align$$
\lcolon S^+(z)S(z)\rcolon
&=1,
\quad
\lcolon S^+(z)S(-z)\rcolon
=\eta(z),
\label{S+S-res}
\\
\lcolon S(\pm\i z)\tS(z)\rcolon
&=\epsilon(\tq^{\mp1/2}z),
\label{StS-res}
\\
\lcolon S^+(\pm\i z)\tS^+(z)\rcolon
&=\epsilon^+(\tq^{\mp1/2}z),
\label{S+tS+-res}
$$}
Again, adding tildes to (\ref{S+S-res}) one more equation can be found.

Now eqs.~(\ref{screening-commut}) can be proved by moving contours. For example, the relation (\ref{S+S-commut}) is obtained as follows. The product $S^+(z)S_k$ is equal to
$$
S^+(z)S_k=\oint_{C_1}{dz'\over2\pi\i}\,z^{k-1}S^+(z)S(z')
=\oint_{C_1}{dz'\over2\pi\i}\,{z^{k-1}\over1-z^{\prime2}/z^2}\lcolon S^+(z)S(z')\rcolon,
$$
where the contour $C_1$ encircles zero leaving the poles $z'=\pm z$ outside. The other product $-z^2S_{k-2}S^+(z)$ is given by the same integral, but over the contour $C_2$, which encircles zero and the points $z'=\pm z$. Their difference
$$
S^+(z)S_k+z^2S_{k-2}S^+(z)=\oint_{C_1-C_2}{dz'\over2\pi\i}\,{z^{k-1}\over1-z^{\prime2}/z^2}\lcolon S^+(z)S(z')\rcolon.
$$
The integral reduces to taking the sum of residues at the poles $z'=\pm z$ and provides~(\ref{S+S-commut}).

All other commutation relations are obtained in the same manner.

\section{Proof of the main statement of Sect.~\ref{S+-sec}}
\label{Rvec-appendix}

Here we prove that the vector defined in~(\ref{t+mod-bra-def}) is an $\cR$\=/vector subject to the condition~(\ref{t+mod-R-bra-condition}).

First of all, rewrite (\ref{t+mod-bra-def}) as
$$
{}_{1n}\langle\ft^{+(p,n)}_{-k_1,\ldots,-k_M}|
={}_{1-2p,n+2M}\langle1|\prod^{\substack{\curvearrowleft\\p}}_{j=1}\tS_{n+2M-2-j(2s-3)}
\prod^{\substack{\curvearrowright\\M}}_{i=1}S^+_{k_i-2p}.
$$
We have
\Multline$$
{}_{1n}\langle\ft^{+(p,n)}_{-k_1,\ldots,-k_M}|{1\over2}d^+_{-l(2s-1)}
=\sum^p_{k=1}{}_{1-2p,n+2M}\langle1|\prod^{\substack{\curvearrowleft\\p}}_{j=k+1}\tS_{n+2M-2-j(2s-3)}
\\\times
\tS_{n+2M-2-k(2s-3)-l(2s-1)}\prod^{\substack{\curvearrowleft\\k-1}}_{j=1}\tS_{n+2M-2-j(2s-3)}
\prod^{\substack{\curvearrowright\\M}}_{i=1}S^+_{k_i-2p}
\\
=(-1)^{p-k}\sum^p_{k=1}{}_{1-2p,n+2M}\langle1|\tS_{n+2M+2p-2-(k+l)(2s-1)}\prod^{\substack{\curvearrowleft\\p}}_{j=k+1}\tS_{n+2M-4-j(2s-3)}
\\\times
\prod^{\substack{\curvearrowleft\\k-1}}_{j=1}\tS_{n+2M-2-j(2s-3)}\prod^{\substack{\curvearrowright\\M}}_{i=1}S^+_{k_i-2p}.
\label{t+mod-d+}
$$
Here and below we assume $l>0$ and odd. We used~(\ref{d+-tS-commut}) to `absorb'~$d^+_{-l(2s-1)}$. The subscript of the first $\tS$ on the r.h.s.\ reads
$$
n+2M+2p-2-(k+l)(2s-1)=(n+2M-n_{p+1})+(p+1-k-l)(2s-1)-1.
$$
Let us assume the condition $n+2M\le n_{p+1}$ to be satisfied. Then this quantity is negative subject to $l>p-k$, and the corresponding term in~(\ref{t+mod-d+}) vanishes. If $l\le p-k$ let us pull the factor $\tS_{n+2M-4-j(2s-3)}$ with $j=k+l$ to the left. We get the product
$$
\tS_{n+2M+2p-2-(k+l)(2s-1)}\tS_{n+2M+2p-4-(k+l)(2s-1)}=0.
$$
The corresponding term in~(\ref{t+mod-d+}) vanishes as well. Hence, the whole r.h.s.\ of~(\ref{t+mod-d+}) is zero. We proved the first condition of the $\cR$\=/vector.

Now let us prove the second condition. Due to~(\ref{d+-tS-commut}) we rewrite the condition in the form
\eq$$
{}_{1n}\langle\ft^{+(p,n)}_{-k_1,\ldots,-k_M}|\tS_{n-2-l(2s-1)}=0.
\label{t+mod-Rvec-cond2}
$$
If $l>p$, by pulling the mode $\tS_{n-2-l(2s-1)}$ to the very left we get $\langle1|\tS_{n+2M-n_{p+1}+(p+1-l)(2s-1)-1}=0$, since the subscript is negative. If $l\le p$, we pull $\tS_{n-2-l(2s-1)}$ till the factor with $j=l$ we get the product
$$
\tS_{n+2M-2-l(2s-3)}\tS_{n+2M-4-l(2s-3)}=0,
$$
so that the condition~(\ref{t+mod-Rvec-cond2}) is always satisfied. This proves that the vector ${}_{1n}\langle\ft^{+(p,n)}_{-k_1,\ldots,-k_M}|$ is an $\cR$\=/vector if $n+2M\le n_{p+1}$. The `nontriviality' condition $n+2M\ge n_p$ follows from the condition $n+2M+2p-2-j(2s-1)>0$ ($1\le j\le p$), that assumes that the corresponding $\tS$ mode does not kill the bra-vacuum and does not map it to another vacuum. The proof for the ket-vector is quite analogous.

\section{Proof of the continuity equations for the currents (\ref{TTheta-posspin-def}),~(\ref{TTheta-negspin-def})}
\label{continuity-negspin-appendix}

Let us consider the $J$\=/function for the operator~$V^{\ft^+_{-2k}\bar c_{-1}}_{11}\sim\bd T_{2k}$:
\Multline*$$
{}_{11}\langle\ft^+_{-2k}|t(X)|c_{-1}\rangle_{11}
={}_{13}\langle1|S^+_{2k}t(X)|c_{-1}\rangle_{11}
=\tB_1^{-1}{}_{13}\langle1|S^+_{2k}t(X)\tS_{-1}|1\rangle_{31}
\\
=\tB_1^{-1}{}_{13}\langle1|S^+_{2k}t(X)\tSigma|1\rangle_{31}
=\tB_1^{-1}{}_{13}\langle1|\tS_1S^+_{2k-2}t(X)|1\rangle_{31}
\\
={}_{33}\langle c_{-1}|S^+_{2k-2}t(X)|1\rangle_{31}
={}_{31}\langle c_{-1}\ft^+_{2-2k}|t(X)|1\rangle_{31}
={}_{-1,-3}\langle c_{-1}\ft^+_{2-2k}|t(X)|1\rangle_{-1,-3}.
$$
The last equation is proportional to $\d\Theta_{2k-2}(x)$ with the right coefficient (see~\cite{Lashkevich:2013yja} for details), which proves that the currents $T_{2k}$, $\Theta_{2k-2}$ defined in~(\ref{TTheta-posspin-def}) satisfy the continuity equations~(\ref{continuity-posspin}).

The $J$\=/functions of the operator $V^{\cR\,c_{-1}\bar\ft^+_{-2k}}_{13}(x)\sim\d T_{-2k}$ are given by
\Multline*$$
{}_{11}\langle c_{-1}|t(X)|\ft^+_{-2k}\rangle_{11}
+{}_{11}\langle c_{-1}|{1\over\pi}[t'(X),\Sigma]\tSigma|\ft^+_{-2k}\rangle_{33}
\\
=\tB_1^{-1}\left(
  {}_{-1,1}\langle1|\tSigma t(X)S^+_{-2k}|1\rangle_{1,-1}
  +{}_{-1,1}\langle1|{1\over\pi}\tSigma[t'(X),\Sigma]\tSigma S^+_{-2k}|1\rangle_{31}\right)
\\
=\tB_1^{-1}\left(
  {}_{-1,1}\langle1|t(X)S^+_{2-2k}\tSigma|1\rangle_{1,-1}
  +{}_{-1,1}\langle1|{1\over\pi}\tSigma[t'(X),\Sigma]S^+_{2-2k}\tSigma|1\rangle_{31}\right)
\\
={}_{-1,1}\langle1|t(X)|c_{-1}\ft^+_{2-2k}\rangle_{-1,1}
+{}_{-1,1}\langle1|{1\over\pi}\tSigma[t'(X),\Sigma]|c_{-1}\ft^+_{2-2k}\rangle_{13}.
$$
By means of the same transformation as we used while deriving~(\ref{V'hh'-delta}) we obtain that the last term is equal to
$$
-{}_{-1,1}\langle1|[t(X),\epsilon_0]|c_{-1}\ft^+_{2-2k}\rangle_{13}
={}_{1,3}\langle1|t(X)|c_{-1}\ft^+_{2-2k}\rangle_{13}-{}_{-1,1}\langle1|t(X)|c_{-1}\ft^+_{2-2k}\rangle_{-1,1}.
$$
Hence,
$$
{}_{11}\langle c_{-1}|t(X)|\ft^+_{-2k}\rangle_{11}
+{}_{11}\langle c_{-1}|{1\over\pi}[t'(X),\Sigma]\tSigma|\ft^+_{-2k}\rangle_{33}
={}_{1,3}\langle1|t(X)|c_{-1}\ft^+_{2-2k}\rangle_{13}.
$$
The r.h.s.\ is proportional to $\bd\Theta_{2-2k}$, which (after checking the coefficients) proves that the currents defined in~(\ref{TTheta-negspin-def}) satisfy the continuity equations~(\ref{continuity-negspin}).

Note that the same argument can be applied to the operators $V^{\cR\>\ft^+_{1-2k}\bc_{-1}}_{11}$ and $V^{\cR\>c_{-1}\bar\ft^+_{1-2k}}_{11}$, which means that the operators produced by the elements $\ft^+_{1-2k}$ are conserved currents on \emph{odd} spin. Nevertheless, it is not difficult to see that these elements can be expanded as
$$
\ft^+_{1-2k}=\sum^k_{j=1}\ft^+_{2j-2k}h_{1-2j},
$$
where $h_{1-2j}\in\check\cA_{2j-1}$ are constructed of odd level generating elements $c_{1-2l}$ only, which, as we know, correspond to commutators with integrals of motion. Hence, the odd spin integrals of motion are not independent.

\raggedright
\bibliographystyle{mybib}
\bibliography{main}

\begin{thebibliography}{10}

\bibitem{Feigin:2008hs}
B.~Feigin and M.~Lashkevich,
\newblock {\it J.~Phys.} {\bf A42}, 304014 (2009), {\tt arXiv:0812.4776}.
%%CITATION = 0812.4776;%%

\bibitem{Alekseev:2009ik}
O.~Alekseev and M.~Lashkevich,
\newblock {\it JHEP} {\bf 1007}, 095 (2010), {\tt arXiv:0912.5225}.
%%CITATION = 0912.5225;%%

\bibitem{Lashkevich:2013mca}
M.~Lashkevich and Y.~Pugai,
\newblock {\it JHEP} {\bf 1309}, 095 (2013), {\tt arXiv:1305.1674}.
%%CITATION = ARXIV:1305.1674;%%

\bibitem{Lashkevich:2013yja}
M.~Lashkevich and Y.~Pugai,
\newblock {\it Nucl.\ Phys.} {\bf B877}, 538 (2013), {\tt arXiv:1307.0243}.
%%CITATION = ARXIV:1307.0243;%%

\bibitem{Lashkevich:2014rua}
M.~Lashkevich and Y.~Pugai,
\newblock {\it JHEP} {\bf 1412}, 112 (2014), {\tt arXiv:1411.1374}.
%%CITATION = ARXIV:1411.1374;%%

\bibitem{Zamolodchikov:1987jf}
A.~B. Zamolodchikov,
\newblock {\it JETP Lett.} {\bf 46}, 160 (1987).
%%CITATION = JTPLA,46,160;%%

\bibitem{Zamolodchikov:1989zs}
A.~B. Zamolodchikov,
\newblock {\it Adv.\ Stud.\ Pure Math.} {\bf 19}, 641 (1989).
%%CITATION = ASPME,19,641;%%

\bibitem{Belavin:1984vu}
A.~A. Belavin, A.~M. Polyakov, and A.~B. Zamolodchikov,
\newblock {\it Nucl.\ Phys.} {\bf B241}, 333 (1984).
%%CITATION = NUPHA,B241,333;%%

\bibitem{Zamolodchikov:1990bk}
{\relax Al}.~B. Zamolodchikov,
\newblock {\it Nucl.\ Phys.} {\bf B348}, 619 (1991).
%%CITATION = NUPHA,B348,619;%%

\bibitem{Belavin:2005xg}
V.~A. Belavin and O.~V. Miroshnichenko,
\newblock {\it JETP Lett.} {\bf 82}, 679 (2005), {\tt arXiv:hep-th/0511128}.
%%CITATION = HEP-TH/0511128;%%

\bibitem{Fateev:2006js}
V.~A. Fateev, V.~V. Postnikov, and Y.~P. Pugai,
\newblock {\it JETP Lett.} {\bf 83}, 172 (2006), {\tt arXiv:hep-th/0601073}.
%%CITATION = HEP-TH/0601073;%%

\bibitem{Fateev:2009kp}
V.~A. Fateev and Y.~P. Pugai,
\newblock {\it J.~Phys.} {\bf A42}, 304013 (2009), {\tt arXiv:0909.3347}.
%%CITATION = 0909.3347;%%

\bibitem{Smirnov:1992vz}
F.~A. Smirnov,
\newblock {\it Adv.\ Ser.\ Math.\ Phys.} {\bf 14}, 1 (1992).
%%CITATION = 00304,14,1;%%

\bibitem{Belavin:2003pu}
A.~A. Belavin, V.~A. Belavin, A.~V. Litvinov, Y.~P. Pugai, and {\relax Al}.~B.
  Zamolodchikov,
\newblock {\it Nucl.\ Phys.} {\bf B676}, 587 (2004), {\tt
  arXiv:hep-th/0309137}.
%%CITATION = HEP-TH/0309137;%%

\bibitem{Lukyanov:1993pn}
S.~L. Lukyanov,
\newblock {\it Commun.\ Math.\ Phys.} {\bf 167}, 183 (1995), {\tt
  arXiv:hep-th/9307196}.
%%CITATION = HEP-TH/9307196;%%

\bibitem{Lukyanov:1997bp}
S.~L. Lukyanov,
\newblock {\it Mod.\ Phys.\ Lett.} {\bf A12}, 2543 (1997), {\tt
  arXiv:hep-th/9703190}.
%%CITATION = HEP-TH/9703190;%%

\bibitem{Dotsenko:1984nm}
{\relax Vl}.~S. Dotsenko and V.~A. Fateev,
\newblock {\it Nucl.\ Phys.} {\bf B240}, 312 (1984).
%%CITATION = NUPHA,B240,312;%%

\bibitem{Cardy:1990pc}
J.~L. Cardy and G.~Mussardo,
\newblock {\it Nucl.\ Phys.} {\bf B340}, 387 (1990).
%%CITATION = NUPHA,B340,387;%%

\bibitem{Babelon:1996sk}
O.~Babelon, D.~Bernard, and F.~A. Smirnov,
\newblock {\it Commun.\ Math.\ Phys.} {\bf 186}, 601 (1997), {\tt
  arXiv:hep-th/9606068}.
%%CITATION = HEP-TH/9606068;%%

\bibitem{Babujian:2002fi}
H.~Babujian and M.~Karowski,
\newblock {\it J.~Phys.} {\bf A35}, 9081 (2002), {\tt arXiv:hep-th/0204097}.
%%CITATION = HEP-TH/0204097;%%

\bibitem{Babujian:2003za}
H.~Babujian and M.~Karowski,
\newblock {\it Phys.\ Lett.} {\bf B575}, 144 (2003), {\tt
  arXiv:hep-th/0309018}.
%%CITATION = HEP-TH/0309018;%%

\bibitem{Delfino:2004vc}
G.~Delfino and G.~Niccoli,
\newblock {\it Nucl.\ Phys.} {\bf B707}, 381 (2005), {\tt
  arXiv:hep-th/0407142}.
%%CITATION = HEP-TH/0407142;%%

\bibitem{Delfino:2005wi}
G.~Delfino and G.~Niccoli,
\newblock {\it J.~Stat.\ Mech.} {\bf 0504}, P04004 (2005), {\tt
  arXiv:hep-th/0501173}.
%%CITATION = HEP-TH/0501173;%%

\bibitem{Delfino:2006te}
G.~Delfino and G.~Niccoli,
\newblock {\it JHEP} {\bf 0605}, 035 (2006), {\tt arXiv:hep-th/0602223}.
%%CITATION = HEP-TH/0602223;%%

\bibitem{Koubek:1993ke}
A.~Koubek and G.~Mussardo,
\newblock {\it Phys.\ Lett.} {\bf B311}, 193 (1993), {\tt
  arXiv:hep-th/9306044}.
%%CITATION = HEP-TH/9306044;%%

\bibitem{Smirnov:1995jp}
F.~A. Smirnov,
\newblock {\it Nucl.\ Phys.} {\bf B453}, 807 (1995), {\tt
  arXiv:hep-th/9501059}.
%%CITATION = HEP-TH/9501059;%%

\bibitem{Jimbo:2003ne}
M.~Jimbo, T.~Miwa, E.~Mukhin, and Y.~Takeyama,
\newblock {\it Commun.\ Math.\ Phys.} {\bf 245}, 551 (2004), {\tt
  arXiv:math/0305323}.
%%CITATION = MATH/0305323;%%

\bibitem{Delfino:2007bt}
G.~Delfino and G.~Niccoli,
\newblock {\it Nucl.\ Phys.} {\bf B799}, 364 (2008), {\tt arXiv:0712.2165}.
%%CITATION = ARXIV:0712.2165;%%

\bibitem{Boos:2006mq}
H.~Boos, M.~Jimbo, T.~Miwa, F.~Smirnov, and Y.~Takeyama,
\newblock {\it Commun.\ Math.\ Phys.} {\bf 272}, 263 (2007), {\tt
  arXiv:hep-th/0606280}.
%%CITATION = HEP-TH/0606280;%%

\bibitem{Jimbo:2010jv}
M.~Jimbo, T.~Miwa, and F.~Smirnov,
\newblock {\it Lett.\ Math.\ Phys.} {\bf 96}, 325 (2011), {\tt
  arXiv:1007.0556}.
%%CITATION = ARXIV:1007.0556;%%

\bibitem{Jimbo:2011bc}
M.~Jimbo, T.~Miwa, and F.~Smirnov,
\newblock {\it Nucl.\ Phys.} {\bf B852}, 390 (2011), {\tt arXiv:1105.6209}.
%%CITATION = ARXIV:1105.6209;%%

\bibitem{LeClair:1989wy}
A.~LeClair,
\newblock {\it Phys.\ Lett.} {\bf B230}, 103 (1989).
%%CITATION = PHLTA,B230,103;%%

\bibitem{Smirnov:1990vm}
F.~A. Smirnov,
\newblock {\it Nucl.\ Phys.} {\bf B337}, 156 (1990).
%%CITATION = NUPHA,B337,156;%%

\bibitem{Reshetikhin:1989qg}
N.~Reshetikhin and F.~Smirnov,
\newblock {\it Commun.\ Math.\ Phys.} {\bf 131}, 157 (1990).
%%CITATION = CMPHA,131,157;%%

\bibitem{Eguchi:1989dq}
T.~Eguchi and S.-K. Yang,
\newblock {\it Phys.\ Lett.} {\bf B235}, 282 (1990).
%%CITATION = PHLTA,B235,282;%%

\bibitem{Lukyanov:1996jj}
S.~L. Lukyanov and A.~B. Zamolodchikov,
\newblock {\it Nucl.\ Phys.} {\bf B493}, 571 (1997), {\tt
  arXiv:hep-th/9611238}.
%%CITATION = HEP-TH/9611238;%%

\bibitem{Zamolodchikov:2004ce}
A.~B. Zamolodchikov,
\newblock {\tt arXiv:hep-th/0401146}.
%%CITATION = HEP-TH/0401146;%%

\bibitem{Fateev:1997yg}
V.~Fateev, S.~L. Lukyanov, A.~B. Zamolodchikov, and {\relax Al}.~B.
  Zamolodchikov,
\newblock {\it Nucl.\ Phys.} {\bf B516}, 652 (1998), {\tt
  arXiv:hep-th/9709034}.
%%CITATION = HEP-TH/9709034;%%

\bibitem{Alekseev:2011my}
O.~Alekseev,
\newblock {\it JETP Lett.} {\bf 95}, 201 (2012), {\tt arXiv:1106.4758}.
%%CITATION = ARXIV:1106.4758;%%

\bibitem{Alekseev:2012jd}
O.~Alekseev,
\newblock {\it JHEP} {\bf 1307}, 112 (2013), {\tt arXiv:1210.2818}.
%%CITATION = ARXIV:1210.2818;%%

\end{thebibliography}

\end{document}